\documentclass{article}
\usepackage{titlesec}
\usepackage{authblk}
\usepackage{multirow}
\usepackage{setspace}
\usepackage{tabularx,booktabs}
\newcolumntype{Y}{>{\raggedright\arraybackslash}X}

\usepackage[english]{babel}
\usepackage[a4paper,top=2.5cm,bottom=2.5cm,left=2.5cm,right=2.5cm,marginparwidth=2cm]{geometry}

\usepackage{amsmath}
\usepackage{graphicx}
\usepackage{float}
\usepackage{siunitx}
\graphicspath{ {./Figures/} }
\usepackage[colorlinks=true, allcolors=blue]{hyperref}

\usepackage{tikz}
\usepackage{pifont}
\usepackage{cleveref}
\usepackage{cite}
\usepackage{array}
\usepackage{booktabs} 

\usepackage{listings}
\usepackage{xcolor} % for setting colors

\renewenvironment{abstract}{%
    \noindent\begin{center}\bfseries\abstractname\end{center}%
    \quotation
}{%
    \endquotation
}

\title{A unified fluid model for nonthermal plasmas and reacting flows}
\author{{Xiao Shao$^{*}$, Deanna A. Lacoste, Hong G. Im }\\
    {\footnotesize Clean Energy Research Platform, Physical Science and Engineering Division, King Abdullah University of Science and Technology (KAUST), Thuwal 23955-6900, Saudi Arabia} \\
    {\footnotesize $^*$E-mail: xiao.shao@kaust.edu.sa}
}

\begin{document}
\date{} % Remove the date
\onehalfspacing

\vspace{40pt}
\maketitle
\begin{abstract}
This work presents a unified fluid modeling framework for reacting flows coupled with nonthermal plasmas (NTPs). Building upon the gas-plasma kinetics solver, ChemPlasKin, and the CFD library, OpenFOAM, the integrated solver, \textit{reactPlasFOAM}, allows simulation of fully coupled plasma-combustion systems with versatility and high performance. By simplifying the governing equations according to the dominant physical phenomena at each stage, the solver seamlessly switches between four operating modes: streamer, spark, reacting flow, and ionic wind, using coherent data structures. Unlike conventional streamer solvers that rely on pre-tabulated or fitted electron transport properties and reaction rates as functions of the reduced electric field or electron temperature, our approach solves the electron Boltzmann equation (EBE) on the fly to update the electron energy distribution function (EEDF) at the cell level. This enables a high-fidelity representation of evolving plasma chemistry and dynamics by capturing temporal and spatial variations in mixture composition and temperature. To improve computational efficiency for this multiscale, multiphysics system, we employ adaptive mesh refinement (AMR) in the plasma channel, dynamic load balancing for parallelization, and time-step subcycling for fast and slow transport processes. The solver is first verified against six established plasma codes for positive-streamer simulations and benchmarked against Cantera for a freely propagating hydrogen flame, then applied to three cases: (1) spark discharge in airflow; (2) streamer propagation in a premixed flame; and (3) flame dynamics under non-breakdown electric fields. These applications validate the model's ability to predict NTP properties such as fast heating and radical production and demonstrate its potential to reveal two-way coupling between plasma and combustion.

\vspace{10pt}
\noindent {\em Keywords:} plasma-assisted combustion (PAC); streamer propagation; electric field; nanosecond repetitively pulsed discharge; computational fluid dynamics (CFD)

\end{abstract}

\vspace{40pt}

\newpage
\section{Introduction} \label{intro}
% Introduce NTPs, PAC and the inverse problem
Nonthermal plasmas (NTPs) are partially ionized gases in which electrons acquire a few eV of energy, while ions and neutrals remain near the bulk‐gas temperature. The resulting non-equilibrium (\(T_e \gg T_i \simeq T_n\) where subscripts $e,i,n$ stand for electrons, ions, and neutrals) enables highly selective chemistry without appreciable gas heating.  This property underpins a broad spectrum of technologies, including water treatment \cite{takeuchi_review_2020}, aerodynamic flow control \cite{corke_single_2009}, plasma medicine \cite{schlegel_plasma_2013}, and plasma-assisted combustion (PAC) \cite{starikovskiy_plasma-assisted_2013} which is of specific interest in this study.  Nanosecond discharges have been shown to enhance ignition, extend lean flammable limits, and stabilize flames under engine-relevant conditions \cite{ju_plasma_2015, lacoste_flames_2023, choe_plasma_2021}.  While most studies were concerned about  the effect of plasma on combustion, the reverse effect has also been examined recently. Pavan \emph{et al.}\ demonstrated that thermo-acoustic oscillations can modulate streamer morphology and energy deposition, revealing the intricate two-way coupling between discharge and flame \cite{pavan_dynamic_2024}.

% Introduce simulation aspects: importance and methods.
Numerical simulations are crucial for studying plasmas in reacting flows, since accurate and sufficiently resolved experimental measurements are challenging, especially for streamer discharges whose characteristic timescales are in nanoseconds. Streamers are filamentary discharges that occur as precursors to other modes, and their simulation with detailed kinetic mechanisms is becoming more prevalent in the NTP community \cite{marskar_3d_2020, zhang_numerical_2020, ren_critical_2023, wang_quantitative_2023, guo_computational_2023}. However, these applications are limited to quiescent, homogeneous mixtures and neglect thermal-chemical coupling. Consequently, the combustion modeling community have long been working to bridge NTPs and reacting flows. In the simplest physical dimension, a zero-dimensional (0D) model typically focus on detailed gas–plasma kinetics and their interactive chemical and energy transfer pathways. It has been widely used to study plasma-assisted ignition and fuel reforming \cite{mao_effects_2019, bang_kinetic_2023}. To determine the electron-impact reaction rates and electron transport properties, the electron energy distribution function (EEDF) is needed. Since EEDF in NTPs does not follow an equilibrium Maxwellian distribution, it is either directly by solving the electron Boltzmann equation (EBE) on the fly, or prescribed by a given functional form. A comprehensive summary can be found in \cite{shao_chemplaskin_2024}. 

To describe complex interactions between plasma and fluid dynamics, multi-dimensional (multiD) models are used but  details in electron-plasma chemistry and transport cannot be fully included due to the excessive computational cost, and some simplifications are necessary. A common approach uses the phenomenological model  \cite{castela_direct_2016, casey_simulations_2017}, in which the discharge energy is split into several channels such as gas heating and radical production, with the specific breakdown of different channels given by empirical constants. The phenomenological model has been used as a cost-effective approach to capture the physical characteristics of nanosecond-pulsed discharges \cite{bechane_numerical_2021, barleon_large-eddy_2023, taneja_large_2024, shao_computational_2024-1}.
% multiD: from phenomenological to full coupling
Since the discharge morphology and electron-energy pathway are prescribed \emph{a priori}, however, the phenomenological model is limited to quasi-steady NTP behavior in simple, well-characterized geometries, such as nanosecond repetitively pulsed (NRP) pin-pin discharges in air. To extend the model applicability to more complex configurations and rapid transient phenomena, multiD models need to describe the fully coupled physical and chemical processes associated with the plasma dynamics and reactive flows. 

The modeling hierarchy outlined above is summarized in Table~\ref{tab:model_comparison}. One-dimensional (1D) formulations are not treated here but they can serve as a  middle ground to incorporate simplified transport effects into complex chemistry. Under most practical PAC conditions at atmospheric pressure or higher, the ionization coefficient is so large that the ionization front propagates as a streamer or even collapses into multiple filamentary channels, violating the planar symmetry assumed by 1D solvers. Such models remain valuable only in special cases where the plasma can be kept laterally uniform, for example in low-pressure Townsend/glow discharges or in micro-gap devices whose dimensions suppress transverse instabilities.

% ------------------------------------------------------------------
% Table 1 – comparison of modelling strategies
\begin{table}[htbp]
  \centering
  \scriptsize            
  \setlength{\tabcolsep}{5pt}
  \renewcommand{\arraystretch}{1.5}
  \caption{Overview of modeling strategies for nonthermal plasmas in reacting flows.}
  \label{tab:model_comparison}
  \begin{tabularx}{\linewidth}{@{} l l Y Y Y @{}}
    \toprule
    \textbf{Dimension} & \textbf{Plasma model} & \textbf{Pros} & \textbf{Cons} & \textbf{Applications} \\ \midrule
    % --------------------------------------------------------------
    0D & Kinetics &
      {-- Low cost\newline
       -- Detailed chemistry} &
      {-- Homogeneous\newline
       -- No transport\newline
       -- Relies on experimental \(E\)-field and \(n_e\)} &
      {-- Reaction-mechanism development\newline
       -- Chemistry \& energy-pathway analysis} \\[0.6em]
    % --------------------------------------------------------------
    multiD & Phenomenological &
      {-- Low cost\newline
       -- Easy implementation} &
      {-- Predetermined discharges\newline
       -- No plasma dynamics\newline
       -- Narrow operating envelope} &
      {-- Reacting flows actuated by quasi-steady discharges with simple structures} \\[0.6em]
      & {Kinetics \& dynamics} &
      {-- High fidelity\newline
       -- Two-way coupling} &
      {-- Complex implementation\newline
       -- High cost} &
      {-- Plasma–flame interactions\newline
       -- Benchmark for reduced-order models} \\
    \bottomrule
  \end{tabularx}
\end{table}

% Sacarcity of high fidelity multiD solvers
The numerical complexity and high computational cost associated with fully resolved plasma models pose daunting challenges in code development. Although some plasma codes include neutral gas dynamics to describe the thermally induced flow, the combustion chemistry and chemical heating are missing \cite{unfer_modeling_2010, zhu_nanosecond_2017, zhang_numerical_2024}. Streamer simulations in combusting environments are rare. Mao \emph{et al.}\ coupled the 2D plasma solver PASSKEy to the reacting-flow code ASURF+ \cite{zhu_nanosecond_2017, chen_theoretical_2007, mao_modeling_2022}, whereas Barleon \emph{et al.}\ linked the AVIP plasma code to the combustion solver AVBP \cite{cheng_avip_2022, schonfeld_steady_1999, barleon_investigation_2023}. The resulting PAC frameworks enable high-fidelity simulations of plasma-assisted ignition. However, both solvers rely on EEDF pre-calculated with BOLSIG\texttt{+} under the local-field approximation (LFA) \cite{hagelaar_solving_2005, grubert_why_2009}, thus restricting their applicability to homogeneous mixtures and ignition problems.

\begin{figure}[H]
\centering
\includegraphics[width=0.95\linewidth]{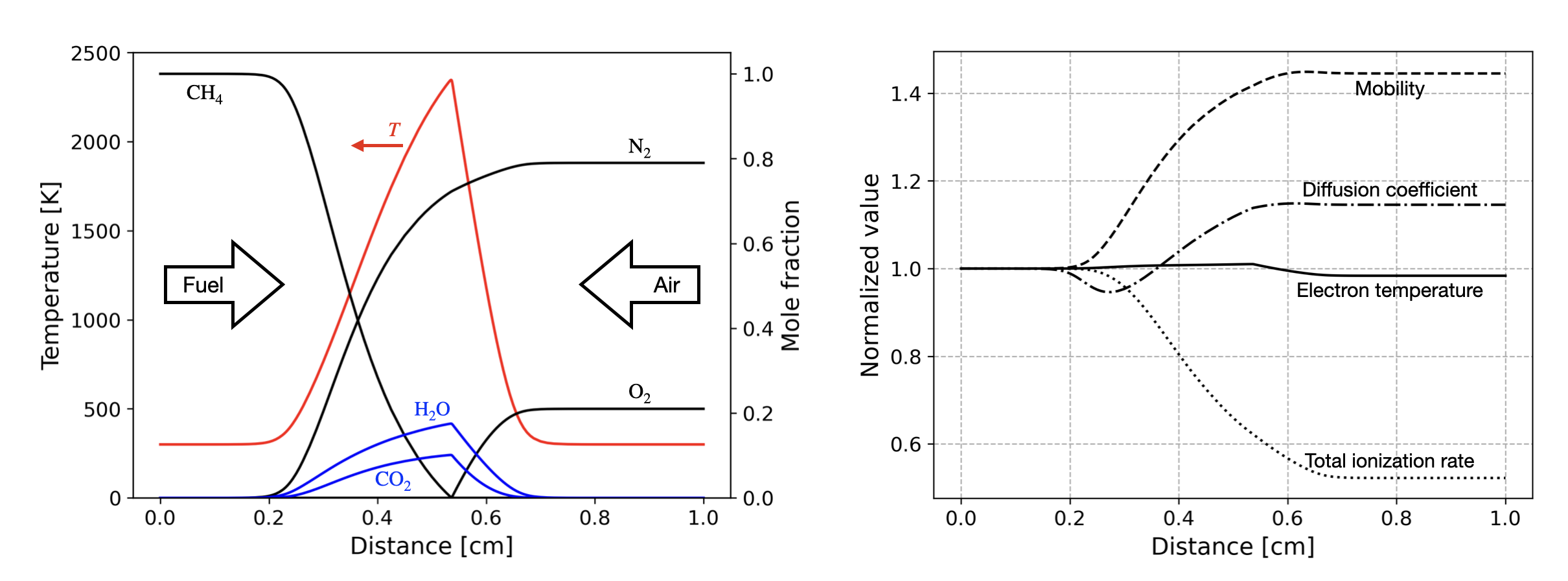}
\caption{\label{fig:flamOnPlasma} Electron properties across a 1-D stoichiometric methane/air diffusion flame at a constant reduced field of 250 Td. Left: flame structure from Cantera. Right: normalized electron temperature, mobility, diffusion coefficient, and ionization rate from CppBOLOS. Cross-section data: LXCat \cite{pancheshnyi_lxcat_2012}.}
\end{figure}

% Highlight the gap in non-homegenous mixtures
To illustrate the influence of mixture composition on NTP electron properties, Figure \ref{fig:flamOnPlasma} shows the electron temperature, mobility, diffusion coefficient, and ionization rate across a one-dimensional (1D) counter-flow methane/air diffusion flame subjected to a constant reduced electric field of 250\,Td (1\,Td = $10^{-21}\,\mathrm{V\,m^2}$).  The computational results were obtained with Cantera \cite{goodwin_cantera_2022} using a one-step mechanism that contains only the five dominant species ($\mathrm{CH_4,\;N_2,\;O_2,\;H_2O,\;CO_2}$). The EBE was solved by CppBOLOS, a C\texttt{++} counterpart to BOLSIG\texttt{+} \cite{shao_chemplaskin_2024}. Although the electron temperature varies little across the flame, the electron mobility and total ionization rate change by more than 40\% between the fuel and air sides. Because these properties are insensitive to the gas temperature, which is an order of magnitude lower than \(T_e\), the variation is driven primarily by composition. Accurate NTP modeling in reacting flows therefore requires accounting for mixture dependence; conventional pre-calculation strategies cannot readily provide high-dimensional lookup tables or curve fits as functions of mixture fraction.

% Goal of this work
The present work aims to advance fully coupled multiD models, address the gap associated with heterogeneous mixtures, and deliver a high-performance implementation. A continuum fluid plasma model is adopted instead of a particle-in-cell/Monte Carlo collision (PIC/MCC) approach because it offers a favorable compromise between space-time resolution and computational cost \cite{alves_foundations_2018}. While LFA is retained, the EBE is solved on each grid cell \emph{on the fly}, allowing local mixture composition to influence electron properties. This plasma model choice facilitates a unified framework for NTPs in reacting flows, because the transport equations for charged and neutral species share the same convection–diffusion–reaction form and can be discretized and solved efficiently within modern CFD libraries. Rather than switching between two stand-alone fluid solvers, the framework developed here yields a single code base that eliminates the data-exchange layer and treats the multiphysical system consistently and efficiently. The strategy builds on our earlier work, ChemPlasKin, a 0D gas-plasma kinetics solver that handles all reactions and thermal effects in a fully coupled ODE system \cite{shao_chemplaskin_2024}. The unified model is implemented by integrating ChemPlasKin into OpenFOAM \cite{weller_tensorial_1998}, resulting in a new solver named \textit{reactPlasFoam}.

% Paper structure
The remainder of this paper is organized as follows.  
Sec.~\ref{methodology} details the governing equations, their simplifications for distinct physical regimes, and the code development.  
Sec.~\ref{benchmark} benchmarks reactPlasFoam against six other streamer solvers in a standardized positive-streamer test case and tests its improved transport model for reacting flows in a 1D freely propagating hydrogen flame against Cantera.  
Sec.~\ref{applications} presents three application studies that showcase the solver’s versatility, accuracy and performance.

\newpage
\section{Formulation} \label{methodology}
\subsection{Governing equations} \label{governing_equations}
The unified fluid model for NTPs coupled with reacting flows consists of conservation equations for mass, momentum, species, and sensible enthalpy, plus Poisson’s equation for the electric potential:
\begin{align}
% Mass Conservation
\frac{\partial \rho}{\partial t} + \nabla \cdot(\rho \mathbf{v}) &= 0 
\label{eq:mass} \,, \\
% 
% Momentum Conservation
\frac{\partial \rho \mathbf{v}}{\partial t} + \nabla \cdot(\rho \mathbf{v v} + p \mathbf{I} - \boldsymbol{\tau}) &= \sum_{i=1}^N q_i e \mathbf{E} n_i 
\label{eq:momentum} \,, \\
%
% Species Mass Conservation
\frac{\partial \rho Y_i}{\partial t} + \nabla \cdot\left[\rho\left(\mathbf{v} + \mathbf{V}_i\right) Y_i\right] &= W_i \dot{\omega}_i + \frac{ W_i S_{\text{ph},i} }{N_\text{A}} 
\label{eq:species} \,, \\
%
% Sensible Enthalpy Conservation
\frac{\partial \rho h^s}{\partial t} + \nabla \cdot\left(\rho \mathbf{v} h^s + \mathbf{q} \right) &= \frac{\partial p}{\partial t} + \mathbf{v} \cdot \nabla p +\dot{Q_\text{r}} + \dot{E}_\text{p}
\label{eq:enthalpy} \,, \\
%
% Poisson's Equation (Electrostatics)
\nabla \cdot \left(\epsilon_r \epsilon_0 \nabla \phi \right) &= -n_c = - \sum_{i=1}^N q_i e n_i \,, \qquad \mathbf{E} = - \nabla \phi \,.
\label{eq:Poisson} 
\end{align}

Here $t$ is the time, $\rho$ the density, $\mathbf{v}$ the bulk flow velocity, $p$ the pressure, $Y_i$ the mass fraction of species $i$ $(i=1,\dots,N)$, and $h^s$ the sensible enthalpy. The electro-hydrodynamic term on the right-hand side of Eq.~(\ref{eq:momentum}) involves the charge number $q_i$, the elementary charge $e = 1.602 \times 10^{-19}\,\mathrm{C}$, the electric field $\mathbf{E}$, and the number density $n_i$ of species $i$. Symbols $\mathbf{I}$ and $\boldsymbol{\tau}$ denote the identity and viscous-stress tensors, respectively. The electric potential $\phi$ is obtained by including the space-charge density $n_c$; $\epsilon_0$ is the vacuum permittivity and $\epsilon_r$ the relative permittivity (dielectric constant).

For neutral species, $\mathbf{V}_i$ simply represents the diffusion velocity. For charged species, the drift term induced by the electric field must be added, giving
\begin{equation}
\mathbf{V}_i = -D_i \nabla \ln Y_i + \frac{q_i}{\left|q_i\right|} \mu_i \mathbf{E} \,,
\label{eq:diffusionVel}
\end{equation}
where $D_i$ and $\mu_i$ are diffusion coefficient and electric mobility of species $i$, respectively. Induced magnetic-field effects are neglected, as justified in \cite{celestin_study_2008}.

In Eq.~(\ref{eq:species}), $W_i$ and $\dot{\omega}_i$ are the molar mass and molar production rate of species $i$, respectively. $S_{\text{ph},i}$ is the photo-ionization source term (see Sec.~\ref{streamer_mode}), and $N_\text{A}$ is Avogadro’s number.

In the energy equation (\ref{eq:enthalpy}), viscous heating is neglected for incompressible flows. The heat-flux vector is
\begin{equation}
    \mathbf{q} = \rho \sum_{i=1}^N Y_i h_i^s \mathbf{V}_i - \lambda \nabla T \,,
\label{eq:heatFlux}
\end{equation}
where $\lambda$ is the mixture thermal conductivity and $T$ the temperature. The volumetric energy sources from chemical reactions and plasma energy deposition are calculated by:
\begin{align}
% Chemical Reaction Heat
\dot{Q_\text{r}} &= - \sum_{i=1}^N h_i^0 \dot{\omega}_i
\label{eq:heatReaction} \,, \\
%
% Plasma Energy
\dot{E}_\text{p} &= \sum_{j \in \mathcal{B}} \varepsilon_\text{th}^j r_\text{B}^j n_e n_{[j]} \,,
\label{eq:Ep}
\end{align}
where $h_i^0$ is the enthalpy of formation of species $i$, and $\mathcal{B}$ is the set of relevant electron–neutral collision processes (e.g., excitation, ionization, dissociation).  
For each process $j$, $\varepsilon_\text{th}^j$ is the threshold energy, $r_\text{B}^j$ the rate coefficient, $n_e$ the electron number density, and $n_{[j]}$ the number density of the neutral collider.  
Alternatively, a global approach based on electron kinetic energy may be used:
\begin{equation}
\dot{E}_\text{p} = e n_e \left( \mu_e \mathbf{E} \right) \cdot \mathbf{E} \,,
\label{eq:Ep2}
\end{equation}
where $\mu_e$ is the electron mobility and $(\mu_e \mathbf{E})$ gives the electron drift velocity. In practice, we find that the discrepancy between these two methods is less than 5\% for well-validated reaction kinetics mechanisms. Note that the portion of plasma energy stored in excited states (i.e., not directly contributing to heating) or consumed by chemical reactions is accounted for in $\dot{Q_\text{r}}$ and $\dot{E}_\text{p}$ via the enthalpy difference between the ground and excited states, thus reflecting the non-equilibrium nature while preserving overall energy conservation. For example, in the sequence
\begin{align}
\mathrm{A + e^-} + \varepsilon_\text{th} &\rightarrow \mathrm{A^* + e^-}  \quad \text{(no heat release)} \label{R:A-A*} \,, \\
\mathrm{A^*} &\rightarrow \mathrm{A} + \varepsilon_\text{th}  \quad \text{(exothermic)} \label{R:A*-A}
\end{align}
the electronic excitation of species A absorbs electron energy $\varepsilon_\text{th}$ into A$^*$. By setting $h^0_{\mathrm{A}^*} = h^0_{\mathrm{A}} + \varepsilon_\text{th}$, Eqs.~(\ref{eq:heatReaction}) and (\ref{eq:Ep}) ensure that reaction~(\ref{R:A-A*}) is nonthermal, while the delayed relaxation of A$^*$ becomes a conventional exothermic reaction.

In a weakly ionized plasma, the electron distribution function, $f_e$, deviates significantly from the equilibrium Maxwellian form and necessitates solving the Boltzmann equation in six-dimensional phase space \cite{hagelaar_solving_2005}:
\begin{equation}
    \label{eq:f_e}
    \frac{\partial f_e}{\partial t} + \mathbf{v}_e \cdot \nabla f_e - \frac{e}{m_e} \mathbf{E} \cdot \nabla_v f_e = \left( \frac{\partial f_e}{\partial t} \right)_\text{coll}\,,
\end{equation}
where $\mathbf{v}_e$ is the electron velocity, $m_e$ the electron mass, $\nabla_v$ the velocity‐space gradient operator, and $(\partial f_e / \partial t)_\text{coll}$ the rate of change due to electron–neutral collisions. A drastic simplification is required to solve Eq.~(\ref{eq:f_e}). Using the two-term expansion under stationary, spatially homogeneous conditions, one obtains the electron energy distribution function (EEDF), denoted $F_0$ \cite{hagelaar_solving_2005, alves_foundations_2018}. Reaction rate coefficients for processes in $\mathcal{B}$ and the electron temperature follow from energy integrals over $F_0$:
\begin{align}
r_\text{B}^j &= \sqrt{\tfrac{2e}{m_e}} \int_0^\infty \epsilon \,\sigma_j(\epsilon)\,F_0(\epsilon)\,d\epsilon\,, \quad j \in \mathcal{B}
\label{eq:r_Bj} \,, \\
T_e &= \frac{2}{3}\,\int_0^\infty \epsilon^{3/2}\,F_0(\epsilon)\,d\epsilon
\label{eq:T_e} \,,
\end{align}
where $\sigma_j$ is the cross section for process $j$. The electron mobility and diffusion coefficient are similarly obtained:
\begin{align}
\mu_e N_0 &= -\frac{\gamma}{3}\,\int_0^\infty \frac{\epsilon}{\tilde\sigma_\text{m}(\epsilon)}\,\frac{\partial F_0}{\partial \epsilon}\,d\epsilon
\label{eq:mu_e} \,, \\
D_e N_0 &= \frac{\gamma}{3}\,\int_0^\infty \frac{\epsilon}{\tilde\sigma_\text{m}(\epsilon)}\,F_0(\epsilon)\,d\epsilon
\label{eq:D_e} \,,
\end{align}
where $N_0$ is the gas number density, $\gamma=\sqrt{2e/m_e}$, and $\tilde\sigma_\text{m}$ the effective momentum‐transfer cross section.  
Although our model recomputes the EEDF on-the-fly for changing gas composition, it retains the locality‐in‐time assumption ($\partial f_e/\partial t=0$) of the local-field or local-mean-energy approximations. The limitations of these stationary, spatially homogeneous EBE solutions are beyond this study’s scope.

The governing equations (\ref{eq:mass}--\ref{eq:Poisson}) are compact but computationally expensive when all closure sub-models are included. Compared to conventional reacting‐flow simulations, typically involving only neutral species in their ground states, adding plasma requires tracking numerous excited states and their complex excitation–relaxation kinetics. The sheer number of species in Eq.~(\ref{eq:species}) can therefore increase by several factors. However, when the kinetic role of vibrational levels of species $k$ is negligible, one can lump all vibrational states and their non‐equilibrium energy into a single variable $e_\text{vib}^k$, greatly reducing species and reaction counts. Furthermore, in a multiphysical discharge–flow problem, it is reasonable to focus on the dominant phenomena in each timescale regime and simplify the equations accordingly. Within our unified framework, the solver operates in four modes: streamer, spark, reacting‐flow, and ionic‐wind, each with tailored simplifications detailed in the following subsections.

\subsection{Special cases}

\subsubsection{Streamer model} \label{streamer_mode}
Streamer discharges occur as precursors to other discharges and represent an important stage in building a conductive plasma channel. The streamer propagation speed can typically reach $10^5$–$10^7\,$m/s and have characteristic times of nanoseconds \cite{celestin_study_2008} in most PAC applications. Background fluid dynamics can be assumed ``frozen" over such short intervals, implying constant bulk density and velocity. Consequently, neutral species are immobile and the governing equations become:
\begin{align}
% Electrons
\frac{\partial n_e}{\partial t} + \nabla \cdot\bigl(-\mu_e n_e \mathbf{E} - D_e \nabla n_e\bigr) &= N_\text{A}\,\dot{\omega}_e + S_{\text{ph},e} 
\quad (\text{electrons})
\label{eq:streamer_ne} \,, \\
%
% Ions
\frac{\partial \rho Y_i}{\partial t} + \nabla \cdot\Bigl(\tfrac{q_i}{|q_i|}\,\rho Y_i \mu_i \mathbf{E} - \rho D_i \nabla Y_i\Bigr) &= W_i\,\dot{\omega}_i + \frac{W_i\,S_{\text{ph},i}}{N_\text{A}} 
\quad (\text{ions})
\label{eq:streamer_ions} \,, \\
%
% Neutrals
\frac{d(\rho Y_i)}{dt} &= W_i\,\dot{\omega}_i + \frac{W_i\,S_{\text{ph},i}}{N_\text{A}} 
\quad (\text{neutrals})
\label{eq:streamer_neutral} \,, \\
%
% Enthalpy
\frac{d(\rho h^s)}{dt} &= \dot{Q}_\text{r} + \dot{E}_\text{p} + \sum_{k\in\mathcal{V}}\bigl(-\dot{E}_{\text{vib}}^k + \dot{R}_{\text{VT}}^k\bigr) 
\label{eq:streamer_hs} \,, \\
%
% Vibrational energy
\frac{d(\rho e_{\mathrm{vib}}^k)}{dt} &= \dot{E}_{\mathrm{vib}}^k - \dot{R}_{\mathrm{VT}}^k 
\quad (k\in\mathcal{V})
\label{eq:streamer_evib} \,,
\end{align}
together with Eq.~(\ref{eq:Poisson}) to update the electric field.  For electrons, we solve for number density rather than mass fraction via $n_e=\rho N_\text{A}Y_e/W_e$, since $Y_e$ can be extremely small ($<10^{-15}$), causing unforgivable floating-point errors during streamer propagation. Number density is thus a safer choice following common practice in plasma community. The background fluid velocity $\mathbf{v}$ is negligible compared to the diffusion–drift velocity and is therefore ignored.

A significant amount of electric energy is channeled into vibrational states via $\dot{E}_{\mathrm{vib}}^k$ for species in the set $\mathcal{V}$. This non-equilibrium vibrational energy is released at a slow heating rate $\dot{R}_{\mathrm{VT}}^k$. For air-dominated mixtures, one may take $\mathcal{V}=\{\mathrm{N_2},\,\mathrm{O_2}\}$; further simplification sometimes retains only $\mathrm{N_2}$ vibrational energy. More details appear in the ChemPlasKin paper \cite{shao_chemplaskin_2024}.

The Zhelezniak photoionization model \cite{zhelezniak_photoionization_1982} is adopted, considering $\mathrm{O_2}$ ionization from radiative quenching of excited singlet states of $\mathrm{N_2}$:
\begin{align}
\mathrm{N_2}(b^1\Pi,\,b'^{1}\Sigma_u^+,\,c_3^1\Pi_u,\,o_3^1\Pi_u,\,c_4'^{1}\Sigma_u^+) &\rightarrow \mathrm{N_2(X)} + h\nu \,, \\
\mathrm{O_2} + h\nu &\rightarrow \mathrm{O_2^+} + e^- \,.
\end{align}
The photoionization source is computed via Helmholtz equations after Bourdon \emph{et al.}\ \cite{bourdon_efficient_2007}:
\begin{align}
\nabla^2 S_\text{ph}^j - (\lambda_j\,p_{\mathrm{O_2}})^2\,S_\text{ph}^j &= -A_j\,p_{\mathrm{O_2}}^2\,I \,, \\
S_\text{ph,e} &= \sum_j S_\text{ph}^j \,,
\end{align}
where $\lambda_j$ and $A_j$ are three‐exponential fit coefficients from \cite{bourdon_efficient_2007}, $p_{\mathrm{O_2}}$ the $\mathrm{O_2}$ partial pressure, and $I$ the photon source. This formulation allows an efficient parallel solution of $S_\text{ph}$. To ensure charge and species conservation, $S_{\text{ph},i}=S_{\text{ph,e}}$ for $\mathrm{O_2^+}$, $-S_{\text{ph,e}}$ for $\mathrm{O_2}$, and zero for other species. Following common practice \cite{celestin_study_2008}, $S_\text{ph}$ is updated every ten iterations to reduce cost without significant accuracy loss.

\subsubsection{Spark model} \label{spark_mode}
After the streamer discharge phase and once a conductive plasma channel has been established, the current rises rapidly if the applied voltage is maintained. During spark mode, the dielectric time step characterizing electric‐field screening within the conducting channel drops below $1\times10^{-15}\,$s. To avoid this prohibitive time step, we adopt the simplified approach of Tholin \cite{tholin_numerical_2012}, assuming a fixed spatial shape of the electric field while its magnitude follows the applied voltage. The space‐charge density $n_c$ then evolves to satisfy Eq.~(\ref{eq:Poisson}):
\begin{align}
\frac{d \|\mathbf{E}\|}{dt} &= \frac{\|\mathbf{E}\|}{\phi}\,\frac{d\phi}{dt}\,, \\
\frac{d n_c}{dt} &= \frac{n_c}{\phi}\,\frac{d\phi}{dt}\,.
\label{eq:spark_E}
\end{align}

Assuming only electrons have sufficient mobility to adjust $n_c$, all other species are integrated locally, and the photoionization model becomes unnecessary in spark mode:
\begin{equation}
\frac{d(\rho Y_i)}{dt} = W_i\,\dot{\omega}_i
\quad (\text{non‐electron species})
\label{eq:spark_charge}
\end{equation}

After updating the heavy species, the electron density follows from charge conservation. However, because the conduction current is limited by the external power supply (not modeled here), we enforce an upper bound on $n_e$ to prevent nonphysical growth:
\begin{equation}
n_e = \min{\Bigl(n_+ - n_- - n_c,\; n_e^{\text{old}}\frac{P_\mathrm{exp}}{\iiint_V \dot{E}_\text{p}\,dV}\Bigr)}\,,
\label{eq:spark_ne}
\end{equation}
where $n_+$ and $n_-$ are the total number densities of positive and negative ions, respectively; $n_e^{\text{old}}$ is the electron density before updating; $P_\text{exp}$ the experimentally measured electric power; and $\iiint_V \dot{E}_\text{p}\,dV$ the plasma power in the domain. This approximation does not strictly conserve electrons, but the error is negligible compared to the large electron growth during spark mode.

\subsubsection{Reacting flow model}\label{comb_mode}
For pulsed plasma systems such as NRP discharges, the applied voltage drops to zero during the interpulse period and the external electric field becomes null. However, due to the several-orders-of-magnitude difference in diffusion coefficients between electrons and positive ions, charge separation occurs and generates a self-induced ambipolar electric field, which slows electron motion and accelerates ion motion. This effectively makes electrons and positive ions diffuse together with an ambipolar diffusion coefficient \cite{shohet_plasma_2003}:
\begin{equation}
    D_a = \frac{\mu_e D_+ + \mu_+ D_e}{\mu_e + \mu_+} \,,
\end{equation}
where the subscript ``+'' denotes positive ions. Applying $\mu_e \gg \mu_+$ and Einstein’s relation \(D_i = \mu_i k_B T_i / q_i\) yields
\begin{equation}
    D_a = \frac{D_+ + \frac{\mu_+}{\mu_e}D_e}{1 + \frac{\mu_+}{\mu_e}}
    \approx D_+\Bigl(1 + \frac{\mu_+D_e}{\mu_e D_+}\Bigr)
    = D_+\Bigl(1 + \frac{T_e}{T_+}\Bigr)\,.
\end{equation}
Assuming $T_+ \approx T_e$ during the interpulse period gives $D_a \approx 2\,D_+$ for both electrons and ions, avoiding the need to solve the ambipolar electric field and drift velocities. This simplification significantly relaxes the time step restriction caused by fast electron transport. Computational cost can be further reduced by neglecting the electrohydrodynamic force and using a low--Mach--number approximation. The governing equations therefore reduce to the reacting‐flow form with additional transport equations for non‐equilibrium vibrational energy, encompassing Eq.~(\ref{eq:mass}) and:
\begin{align}
% Momentum Conservation
\frac{\partial \rho \mathbf{v}}{\partial t} + \nabla \cdot(\rho \mathbf{v}\mathbf{v} + p \mathbf{I} - \boldsymbol{\tau}) &= 0
\label{eq:momentum_comb} \,, \\
%
% Species Conservation
\frac{\partial \rho Y_i}{\partial t} + \nabla \cdot\bigl[\rho(\mathbf{v} + \mathbf{V}_i) Y_i\bigr] &= W_i \dot{\omega}_i
\label{eq:species_comb} \,, \\
%
% Enthalpy Conservation
\frac{\partial \rho h^s}{\partial t} + \nabla \cdot(\rho \mathbf{v} h^s + \mathbf{q}) &= \frac{d p}{d t} + \dot{Q}_\text{r} + \sum_{k \in \mathcal{V}}\dot{R}_{\mathrm{VT}}^k
\label{eq:enthalpy_comb} \,,\\
%
% Vibrational Energy
\frac{\partial \rho e_{\mathrm{vib}}^k}{\partial t} + \nabla \cdot(\rho e_{\mathrm{vib}}^k \mathbf{v}) &= \nabla \cdot(\rho D_k \nabla e_{\mathrm{vib}}^k) - \dot{R}_{\mathrm{VT}}^k
\label{eq:evib_comb} \,.
\end{align}

Note that we solve for the mass‐fraction form of electrons in this mode to preserve conservation, since $\rho$ cannot be assumed constant and the $Y_e$ accuracy issue from streamer mode becomes negligible here. The vibrational--translational (V--T) relaxation rate is given by:
\begin{equation}
    \label{eq:R_VT}
    \dot{R}_\text{VT}^k = \frac{e_{\text{vib}}^k}{\tau_\text{VT}^k}\,,
\end{equation}
where $\tau_\text{VT}^k$ is the V–T relaxation timescale (see \cite{shao_chemplaskin_2024}). The diffusion coefficient $D_k$ for vibrational energy is taken as the mass diffusivity of carrier species $k$.

A mixture-averaged model determines the species diffusion coefficient:
\begin{equation}
    D_i = \frac{1 - Y_i}{\sum_{j \neq i} X_j D_{ij}} \,,
\end{equation}
where $D_{ij}$ is the binary diffusion coefficient between species $i$ and $j$, and $X_i$ is the mole fraction. $D_i$ is computed using the Cantera module integrated in the solver.

\subsubsection{Ionic wind model}\label{ionic_mode}
When the duration of the applied voltage is comparable to the hydrodynamic timescale, ions drift under continuous Coulomb force and transfer momentum to neutral gas molecules, producing the so-called ionic wind effect \cite{belan_compared_2015, belhi_direct_2010}. In this mode, the electrohydrodynamic force in Eq.~(\ref{eq:momentum}) can no longer be neglected, and the electric field must be updated every time step. The unified model is inherently suitable for studying non-breakdown cases in which weakly ionized reacting flows are subjected to an electric field. For example, Won \emph{et al.}\ studied the effects of electric fields on the reattachment of lifted flames \cite{won_effect_2007}. In this regime, Eqs.~(\ref{eq:mass}--\ref{eq:Poisson}) are mostly retained (photoionization is omitted) and Eq.~(\ref{eq:Ep2}) is used to compute the input electric energy. More details, including ion reactions and transport properties, are given in the modeling work of Belhi \emph{et al.}\ \cite{belhi_direct_2010}.

A major computational challenge in ionic wind mode is the strict time step $\Delta t$ constraint that persists over a much longer physical duration. Compared to streamer propagation where $\Delta t$ may be as small as $1\times10^{-13}\,$s but the event lasts only nanosecond, $\Delta t$ in ionic wind mode typically spans a few nanoseconds, while the total simulated time must reach on the order of milliseconds to capture the weak, slow electrohydrodynamic effects. To improve time-marching efficiency, we exploit the timescale separation between charges and neutrals by integrating the governing equations in two groups with different time steps.

The main group comprises the conservation equations for mass, momentum, neutral species, and enthalpy. These equations, which represent the bulk flow hydrodynamics, are advanced with $\Delta t_{\text{main}}$. Within each main time step, a sub-group containing charged-species transport and Poisson’s equation is integrated internally with a much smaller time step $\Delta t_{\text{sub}}$. The ionic wind mode and the associated sub-cycling technique are examined in Sec.~\ref{efieldFlame}.

\newpage
\section{Computational development} \label{code_dev}
The code development process for efficient solution of the governing equations involves significant software engineering. Several important numerical and acceleration aspects of reactPlasFoam are described in the following subsections.

\subsection{OpenFOAM and ChemPlasKin}
\begin{figure}[H]
  \centering
  \includegraphics[width=0.8\linewidth]{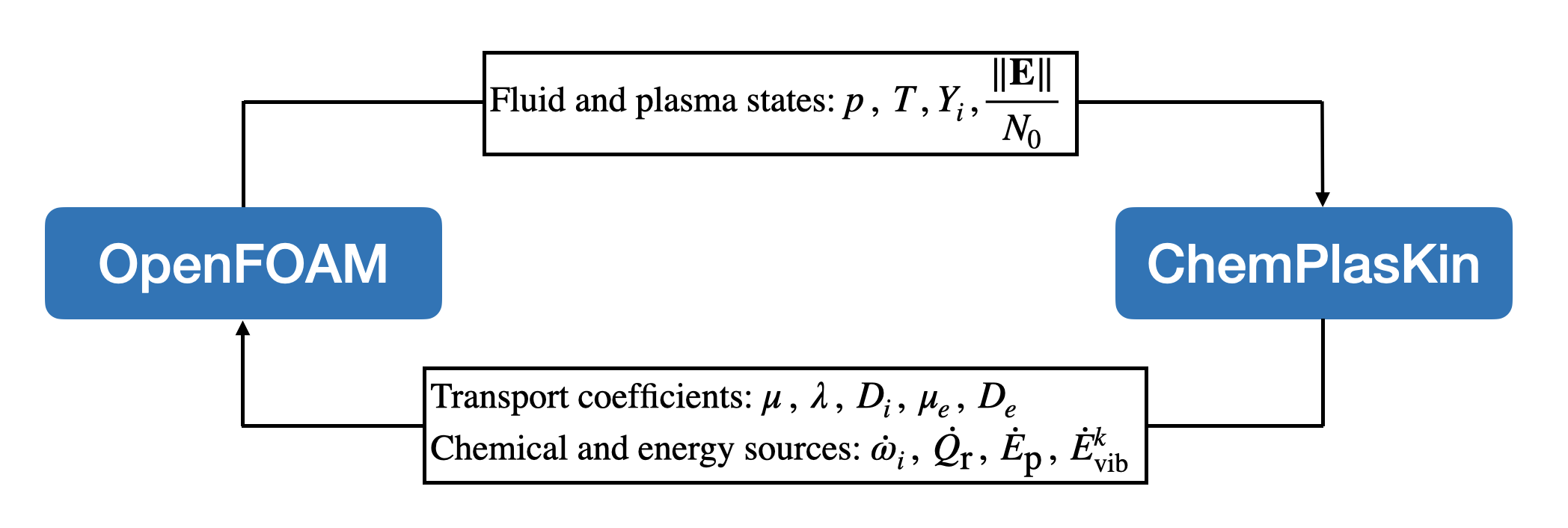}
  \caption{Data communication between OpenFOAM and ChemPlasKin in reactPlasFoam.}
  \label{fig:data_flow}
\end{figure}

ChemPlasKin was released in our previous work, where its potential integration into a CFD solver was mentioned \cite{shao_chemplaskin_2024}. Before that, several groups have accelerated chemical‐source integration and improved the transport model of the original OpenFOAM reacting‐flow solver \textit{reactingFoam} by coupling Cantera with OpenFOAM \cite{li_quasi-direct_2020, yang_reactingfoam-sci_2019, zhou_robust_2022}.  Built on the Cantera library, ChemPlasKin solves plasma and gas kinetics in a unified ODE system and employs the EBE solver CppBOLOS to update the EEDF on-the-fly, making it well suited for incorporation into a C\texttt{++} CFD framework like OpenFOAM.  Figure~\ref{fig:data_flow} illustrates the data communication between OpenFOAM and ChemPlasKin, highlighting the core model-closure functionality of reactPlasFoam. Field states, including pressure, temperature, mixture composition and reduced electric field, are sent to ChemPlasKin. ChemPlasKin returns transport properties and source terms at the cell level. The transport coefficients consist of gas viscosity, thermal conductivity and mass diffusivity, as well as electron mobility and diffusivity. Species production rates and plasma-chemistry energy deposition rates are also computed by ChemPlasKin.

\subsection{Time-step constraints}
Several time-step limitations must be applied to ensure simulation stability and accuracy. The classical Courant--Friedrichs--Lewy (CFL) time step is the primary constraint for hydrodynamics:
\begin{equation}
    \Delta t_\text{CFL} = \frac{\text{CFL}\,\Delta x}{\lVert \mathbf{v}\rVert}\,.
    \label{dt_CFL}
\end{equation}

The stability of diffusion fluxes requires a limiting time step $\Delta t_\text{F}$ based on a Fourier number $F \le 0.25$ for a diffusion coefficient $D$:
\begin{equation}
    \Delta t_\text{F} = \frac{F\,\Delta x^2}{D}\,,
    \label{dt_F}
\end{equation}
a major constraint in streamer mode due to the fine mesh at the streamer head and the large electron diffusion coefficient $D_e$.

Once a conductive plasma channel forms, the dielectric time step becomes most restrictive:
\begin{equation}
    \Delta t_\text{d} = A_\text{d}\frac{\epsilon_0}{\textstyle\sum_i{|q_i n_i \mu_i}|} \simeq A_\text{d} \frac{\epsilon_0}{e n_e \mu_e} \,,
    \label{dt_d}
\end{equation}
where $A_\text{d} = 0.5$ is a suitable approximation. Spark mode is usually triggered when $\Delta t_\text{d}$ falls below $1 \times 10^{-13}$~s. 

In most streamer solvers, a chemical time‐step restriction is also involved:
\begin{equation}
    \Delta t_{\rm chem} = A_{c}\,\min{\Bigl(\frac{\rho\,Y_i}{W_i\lvert\dot{\omega}_i\rvert}\Bigr)}\,,
    \label{dt_c}
\end{equation}
with $A_{c}=0.05$. However, the operator‐splitting technique, a standard method in reacting‐flow simulations that decouples fast chemical kinetics from the slower fluid‐dynamic processes, can also be applied in plasma modeling when $\Delta t_{\rm chem}$ becomes too restrictive. We set the trigger criterion as
\[
    \Delta t_{\rm chem} \le 0.1\,\min\bigl(\Delta t_{\rm CFL},\,\Delta t_{F},\,\Delta t_{d}\bigr).
\]
In that case, the ODE solver CVODE from the SUNDIALS suite \cite{hindmarsh_sundials_2005, gardner_enabling_2022} is invoked via the ChemPlasKin interface to integrate the stiff chemical kinetics within a global time step of $\min\bigl(\Delta t_{\rm CFL},\,\Delta t_{F},\,\Delta t_{d}\bigr)$. When $\Delta t_{\rm chem}$ is comparable to the other restrictions, the time step in streamer mode is chosen as
\begin{equation}
    \Delta t = \min{\bigl(\Delta t_{\rm CFL},\,\Delta t_{F},\,\Delta t_{d},\,\Delta t_{\rm chem}\bigr)}\,.
\end{equation}

Spark mode is computationally cheap due to the simplifications, and a fixed time step \(\Delta t = 2\times10^{-11}\)~s is recommended. Sub-step integration may still be used for chemical kinetics.

\subsection{Adaptive mesh refinement}\label{AMR}
During the streamer propagation stage, the characteristic length scale of a streamer head is
\begin{equation}
    \Delta x = \frac{1}{\alpha}\,,
\end{equation}
where $\alpha$ is the Townsend ionization coefficient, i.e., the number of electrons created by one electron per unit length. The grid size needed to resolve $\Delta x$ is typically a few micrometers. In contrast, reacting‐flow structures in modern CFD studies span from centimeters to a few meters. Thus, using a single uniform mesh for the coupled multiphysical system would be extremely inefficient and impractical.

To address this multiscale meshing challenge, we have implemented the advanced AMR library of Rettenmaier \emph{et al.}\ \cite{rettenmaier_load_2019} in reactPlasFoam. Compared to OpenFOAM’s original AMR which supports only 3D meshes, this library supports both 2D and 2D axisymmetric refinement and includes dynamic load balancing (DLB) for domain re‐partitioning. We use the scalar field $0.5/\alpha$ as the refinement criterion.

Mesh unrefinement offers tremendous benefits when the simulation transitions from the plasma‐dominant stage to the (typically much longer) reacting‐flow stage. In 2D cases, halving the grid size can increase computational speed by roughly a factor of eight. Barleon \cite{barleon_detailed_2022} adopted a dual‐mesh strategy, mapping data via interpolation between fine and coarse meshes for the discharge and interpulse periods, respectively. With AMR, refinement and coarsening occur more efficiently, and data integrity is maintained.  

\subsection{Schemes and solvers}\label{schemes}
The reacting–flow solution follows the PIMPLE (coupling PISO and SIMPLE) algorithm, which operates from incompressible to moderately compressible flows. For spatial discretization of convective terms, a second‐order linear interpolation is used at cell faces, augmented with a flux limiter. For diffusive terms, a second‐order central differencing method is applied with a non‐orthogonality correction.

Time integration of the governing equations is performed using either a first‐order implicit Euler scheme or a second‐order backward differentiation formula (BDF), with the choice guided by the stiffness of the problem and the desired accuracy.

Elliptic equations, including Poisson’s and Helmholtz equations, are solved with the geometric–algebraic multigrid (GAMG) solver, which combines geometric coarsening and algebraic smoothing to efficiently eliminate error components across scales and achieve robust, scalable convergence.

Transport equations for velocity, enthalpy, species mass fractions, and number density are solved using a preconditioned bi-conjugate gradient stabilized (BiCGSTAB) iterative solver, accelerated with incomplete‐factorization preconditioners.

Within the ChemPlasKin module, reaction kinetics are integrated by CVODE from the SUNDIALS suite, which employs adaptive, variable‐order BDF methods designed for stiff ODE systems \cite{hindmarsh_sundials_2005,gardner_enabling_2022}. The EBE solver in ChemPlasKin, CppBOLOS, computes the EEDF using BiCGSTAB with an incomplete‐Cholesky preconditioner provided by the Eigen library \cite{guennebaud_eigen_2010}.

\subsection{Fully coupled EBE solver}
\subsubsection{Challenges}
As discussed in the introduction, the EEDF and electron properties in non-homogeneous reacting flows depend not only on the local reduced electric field but also on the mixture composition. An on-the-fly update of the EEDF in grid cells is necessary for accuracy. Even though the EBE problem is reduced to a 1D linear system via a two-term expansion, iteratively solving it for every cell at each time step can be prohibitively expensive. Indeed, during the initialization phase, over $95\%$ of the computational cost may be consumed by CppBOLOS, which computes $(\mu_e, D_e, r_\mathrm{B}^j)$ for all cells.

A straightforward remedy is to cache previously computed electron properties and update them only in a subset of cells. During streamer propagation, high electric-field variations are typically confined to the streamer boundary region, while the temporal evolution of mixture composition remains relatively small due to the low ionization degree in NTPs and the short duration of streamer propagation. Consequently, we  recompute only in the cells whose reduced electric field has changed beyond user-specified tolerances:
\begin{equation}
  \label{eq:update-set}
  \mathcal{C} := \Bigl\{\, i \Bigm| \delta (E/B)_i\geq
      \max{\bigl(r_{\mathrm{rel}}(E/N)_i,\, r_{\mathrm{abs}}\bigr)}
  \Bigr\} \,,
\end{equation}
where $\delta (E/N)_i$ is the accumulated temporal change of reduced electric field ($\|\mathbf{E}\|/N_0$) at cell $i$, and $r_{\mathrm{rel}}$ and $r_{\mathrm{abs}}$ are the relative and absolute tolerances. After a cell $i\in\mathcal{C}$ is processed, its accumulator $\delta (E/N)_i$ is reset to zero. A full re-initialization is forced at the start of each streamer pulse to capture composition changes after every discharge.

This strategy significantly reduces the EBE solve count. In homogeneous mixtures, a 1D lookup table can cache solutions of $(\mu_e, D_e, r_\mathrm{B}^j)$ for $[0.8\min(E/N), \, 1.2\max(E/N)]$ with sufficient resolution; updated cells fetch interpolated values with minimal overhead, and the table is updated if the field range exceeds its envelope. However, for non-homogeneous mixtures, building a high-dimensional lookup table including mixture composition is impractical. Cells in $\mathcal{C}$ must be submitted to CppBOLOS, causing serious load imbalance in parallelization as each processor faces differing local $\mathcal{C}$ counts, leading to idle processors and reduced scalability.

\subsubsection{Dynamic load balancing (DLB)} \label{DLB}
When solving the EBE for updated cell values $(\mu_e, D_e, r_\mathrm{B}^j)$, we define a \verb|task| as a serialized data structure for cell $i\in\mathcal{C}$ containing the cell index, reduced electric field, and thermochemical states ($T, p, Y_i$). Each task thus comprises $N+3$ double-precision variables plus one integer index.

We propose an algorithm to offload extra tasks from overloaded processors (OPs) to underloaded processors (UPs). This algorithm builds a task allocation map with the following steps:

\begin{enumerate} 
\item Gather all processor task counts and rank them.
\item Compute the rounded average load and capacity in each processor. 
\item The heaviest OP (OP1) offload tasks to the lightest UP (UP1) up to UP1’s capacity. 
\item Continue offloading to UP2, UP3, etc., until OP1 is balanced.
\item Repeat for OP2, OP3, etc., until each OP’s extra tasks are assigned. 
\end{enumerate}

Each processor constructs two complementary maps: \verb|sendersMap| (which OP sends tasks and how many) and \verb|receiversMap| (which UP receives tasks and how many). These two maps mirror one another. Because every processor runs the same allocation procedure without a master node, they all arrive at an identical task allocation map, avoiding communication conflicts.

This “mapping-first” approach offers more flexibility than one-to-one pairing while avoiding the overhead of fully shared pools. Once the map is established, each processor identifies itself as either OP or UP and consults the map to offload or receive tasks. UPs process received tasks and return updated electron properties $(\mu_e, D_e, r_\mathrm{B}^j)$ to the appropriate OPs, which integrate them back into their subdomain cells. Local tasks are handled directly. Further details and pseudo code are provided in the Appendix.

In our Message-Passing Interface (MPI) implementation, we use fixed message tags (e.g., 0, 1, 2, 3) with explicit source/destination ranks to avoid tag conflicts. Non-blocking communication (\verb|MPI_Isend|, \verb|MPI_Irecv|) could overlap offloading with other work but requires careful matching. We found that, as long as each \verb|MPI_Recv| specifies the correct source rank and tag, alignment is preserved even with multiple senders, ensuring stable data exchange without congestion.

\newpage
\section{Code validation} \label{benchmark}
Because no publicly available solver or simulation data exist for NPTs coupled with reacting flows, a complete verification of reactPlasFoam’s functionality cannot be achieved. Alternatively, we tested the code for streamer propagation and reacting flow separately. For the first part of this section, we benchmark our code against six positive-streamer simulation results from Bagheri \emph{et al.}\ \cite{bagheri_comparison_2018}. Four of these used self-developed codes, and two used the commercial COMSOL Multiphysics\textsuperscript{\textregistered}. For the second part, we benchmark reactPlasFoam against Cantera for a 1D freely propagating flame.

\subsection{A simple positive streamer}
\begin{figure}[H]
  \centering
  \includegraphics[width=0.48\linewidth]{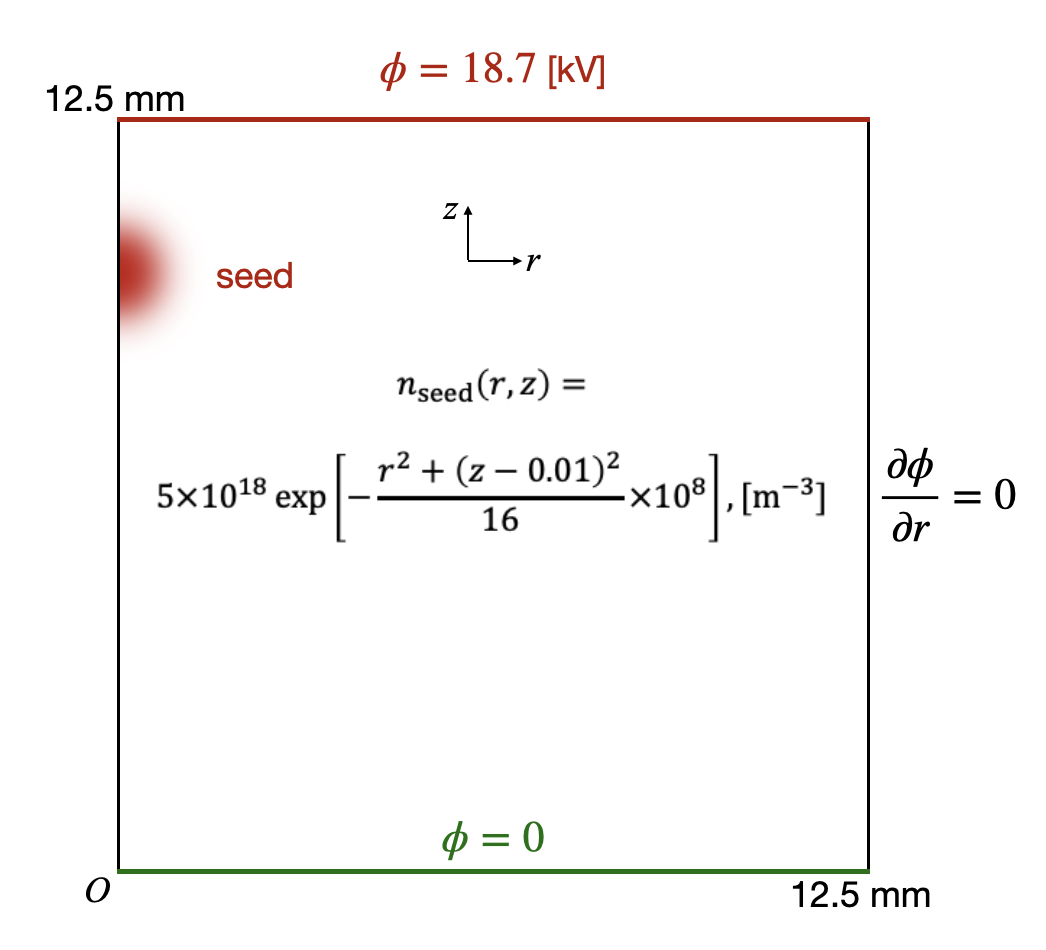}
  \caption{\label{fig:benchmark_config} Schematic of the computational domain used in \cite{bagheri_comparison_2018}. The red region shows the initial seed of positive ions centered at $(r,z)=(0,10)\,$mm.}
\end{figure}

The axisymmetric computational geometry is shown in Figure~\ref{fig:benchmark_config}, where a positive streamer is generated between planar electrodes separated by 12.5~mm. To initialize breakdown, a Gaussian distribution of positive ions is seeded near the anode. The model is highly simplified and considers only electrons and immobile positive ions:
\begin{align}
\frac{\partial n_e}{\partial t} + \nabla \cdot\bigl(-\mu_e\,n_e\,\mathbf{E} - D_e\,\nabla n_e\bigr) &= \bar\alpha\,\mu_e\,\lVert\mathbf{E}\rVert\,n_e + S_{\text{ph},e}\,, \\
\frac{d n_+}{d t} &= \bar\alpha\,\mu_e\,\lVert\mathbf{E}\rVert\,n_e + S_{\text{ph},e}\,,
\end{align}
where $\bar\alpha$ is the effective ionization coefficient, and $\bar\alpha$, $\mu_e$, and $D_e$ are approximated analytically as functions of $\|\mathbf{E}\|$ for dry air at 1~bar and 300~K. Details of this model and numerical setup are in \cite{bagheri_comparison_2018}.

Two cases are tested. The first excludes photoionization and uses a moderate background ionization level of $n_+=n_e=1\times10^{13}\,\mathrm{m^{-3}}$, smoothing charge gradients and reducing streamer-head field strength, thereby relaxing grid and time-step requirements compared to low preionization. The second uses very low initial ionization $n_+=n_e=1\times10^{9}\,\mathrm{m^{-3}}$ and includes photoionization. All simulations were run on eight threads of an AMD EPYC™ 9654 processor (2.4\,GHz base clock). The results are presented as follows.

\subsubsection{Case 1: high pre-ionization}
\begin{figure}[H]
\centering
\includegraphics[width=0.48\linewidth]{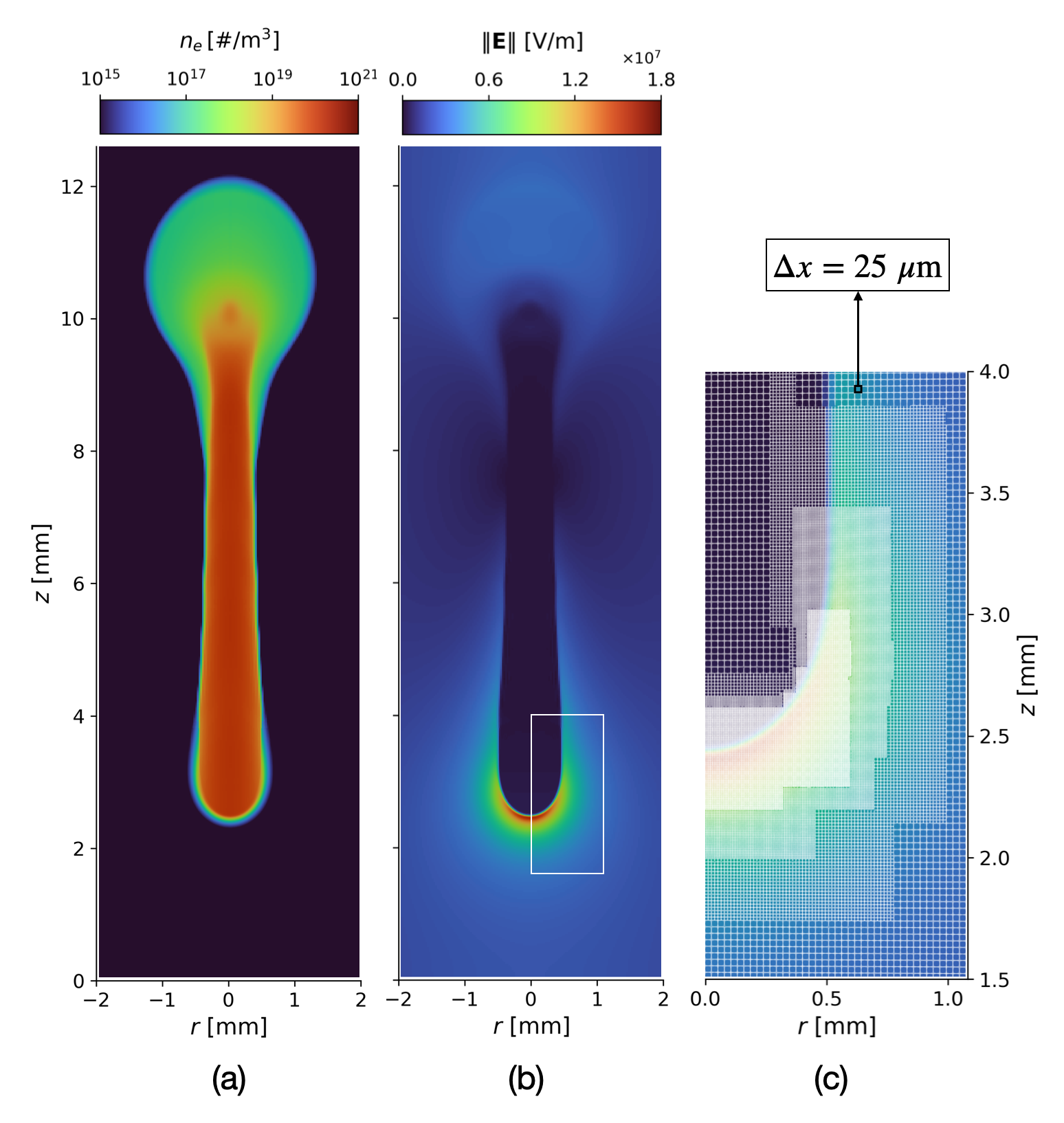}
\caption{\label{fig:case1_map} 2D maps of a positive streamer at $16\,$ns with high background ionization: (a) electron number density; (b) electric field strength; (c) AMR zoom-in at the streamer head with minimum grid size of $3.125\,\mu\mathrm{m}$.}
\end{figure}

Stable streamer propagation up to $16\,$ns was obtained. The electron density and electric field are shown in Figure~\ref{fig:case1_map}. The AMR refinement criterion is set to $\Delta x \le 0.5/\alpha$, with a minimum grid size of $3.125\,\mu\mathrm{m}$. Figure~\ref{fig:case1_map}(c) shows the mesh refinement at the streamer head, where rectangular buffer layers ensure smooth transition of grid splitting. The maximum cell count is about $1.1\times10^5$, close to the CWI group’s results in \cite{bagheri_comparison_2018} using the afivo-streamer code \cite{teunissen_simulating_2017}. reactPlasFoam completes the simulation in $20\,$min, ranking second only to afivo-streamer (5\,min on eight threads at $3.6\,$GHz) in performance. CPU hours reported by the other five groups range from 6 to 90.

\begin{figure}[H]
\centering
\includegraphics[width=0.95\linewidth]{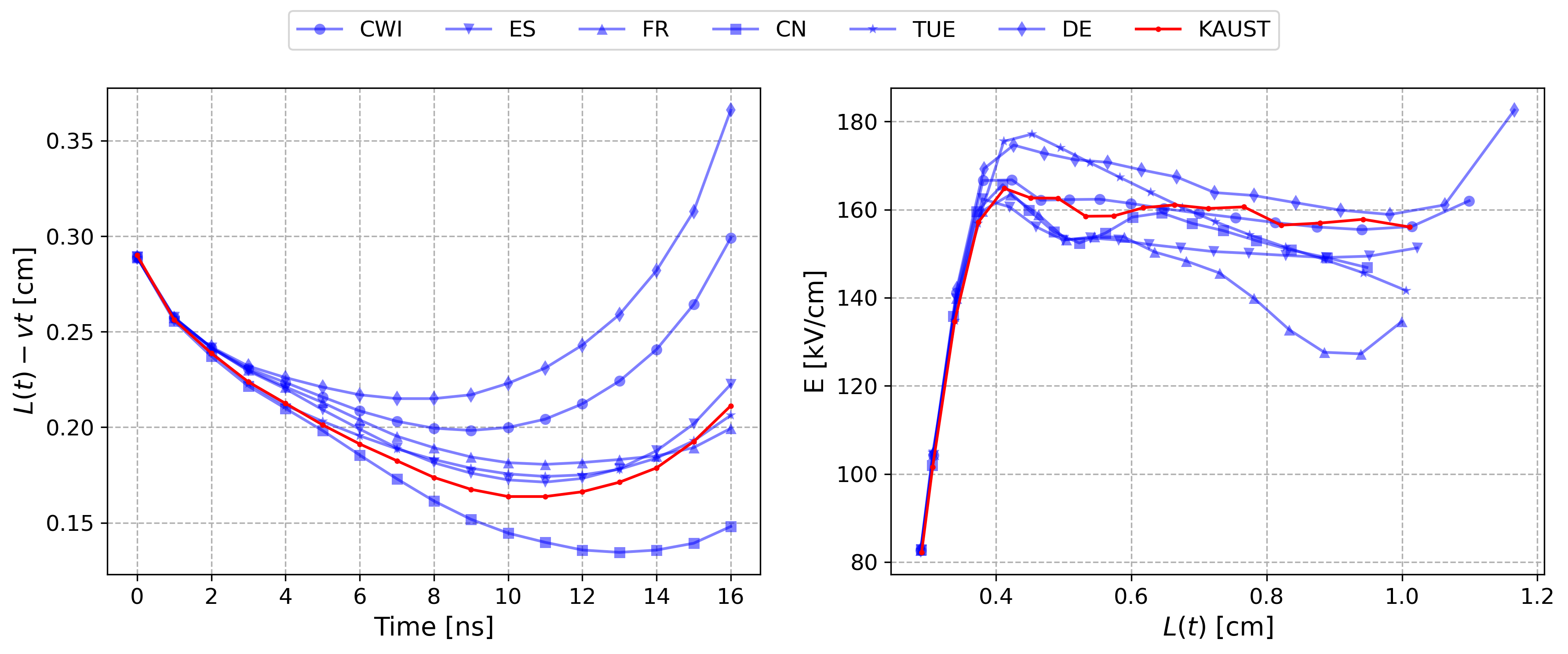}
\caption{\label{fig:case1_solvers} Comparison of reactPlasFoam results (KAUST) for Case~1 with those from \cite{bagheri_comparison_2018}. Left: $L(t) - vt$ versus time ($v=0.05\,\mathrm{cm/ns}$); right: maximum electric field strength versus streamer length.}
\end{figure}

To further evaluate the quantitative accuracy of reactPlasFoam, the streamer length and maximum electric field along the axis are compared with results from other codes (Figure~\ref{fig:case1_solvers}). The streamer length is defined as
\[
  L(t) = 12.5 - z_\mathrm{max}(t)\;\text{[mm]},
\]
where $z_\mathrm{max}(t)$ is the axial location of maximum electric field. The overall agreement of reactPlasFoam with other solvers is good. However, growing differences in streamer location and maximum electric field between some groups underscore the high sensitivity of streamer modeling to numerical variations, though the test case was designed for standardized comparison.

\subsubsection{Case 2: low pre-ionization with photoionization source}
\begin{figure}[H]
  \centering
  \includegraphics[width=0.48\linewidth]{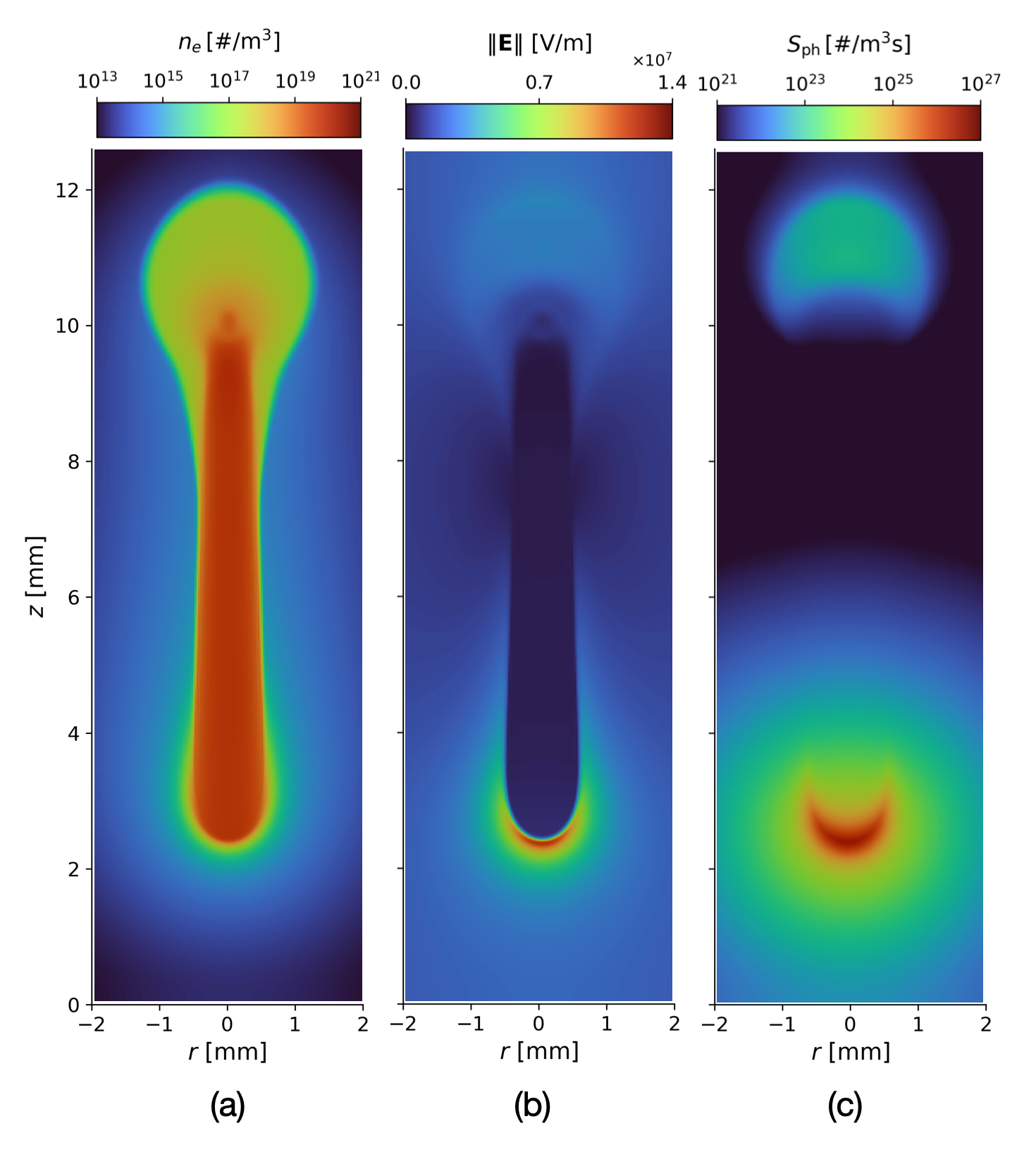}
  \caption{\label{fig:case3_map} 2D maps of a positive streamer at $15\,$ns with low background ionization: (a) electron number density; (b) electric field strength; (c) photoionization source.}
\end{figure}

The second test case uses a low background ionization level and includes the photoionization model of \cite{zhelezniak_photoionization_1982} with the parameterization from \cite{bourdon_efficient_2007}. The streamer state at $15\,$ns is shown in Figure~\ref{fig:case3_map}. The addition of the photoionization source generates extra electrons and significantly reduces the numerical stiffness at the streamer head. Without photoionization, a low pre-ionization condition typically leads to higher gradients of electron density, stronger electric fields, and faster ionization rates.

\begin{figure}[H]
  \centering
  \includegraphics[width=0.95\linewidth]{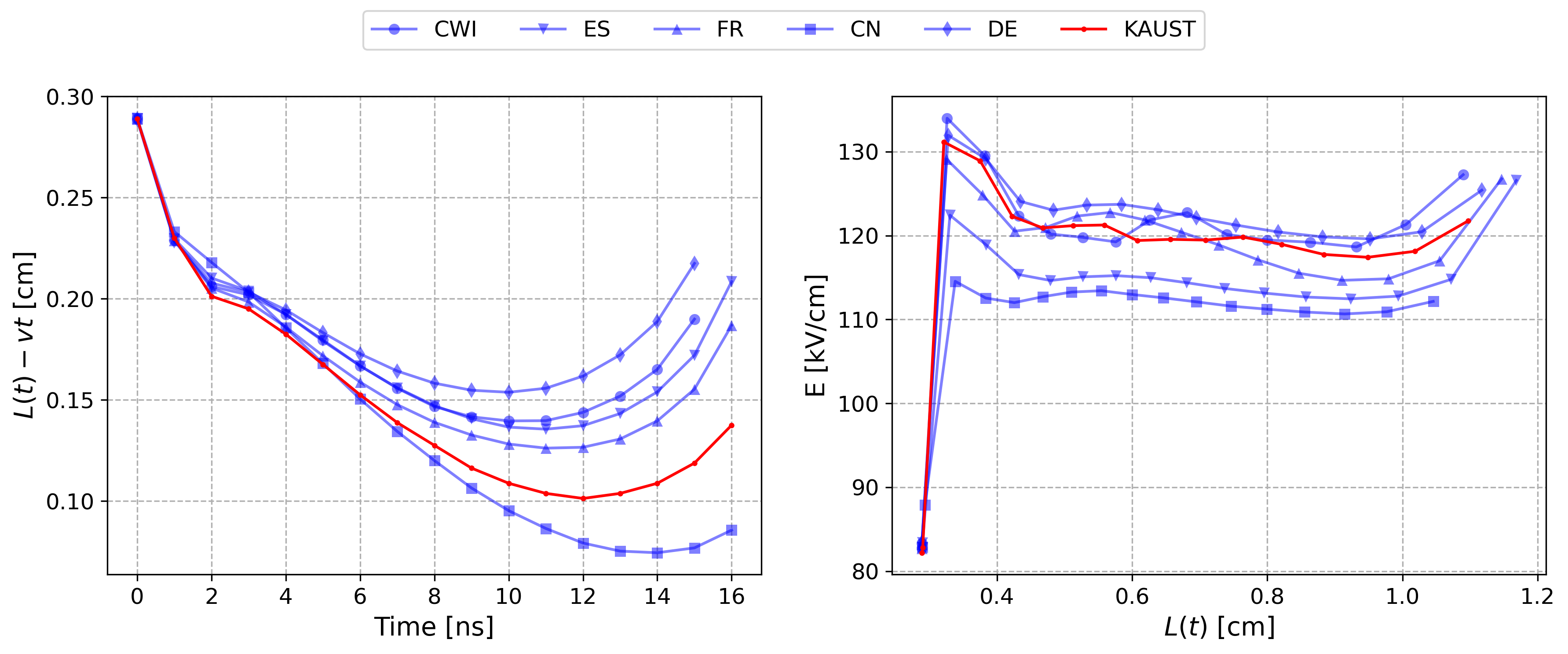}
  \caption{\label{fig:case3_solvers} Comparison of reactPlasFoam results (KAUST) for Case 2 with those from \cite{bagheri_comparison_2018}. Left: $L(t) - vt$ versus time ($v=0.06\,\mathrm{cm/ns}$); right: maximum electric field strength versus streamer length.}
\end{figure}

Once again, the streamer length and maximum electric field are compared in Figure~\ref{fig:case3_solvers}, showing satisfactory performance of reactPlasFoam within the solution envelope.

\subsection{1D freely propagating hydrogen flame}
The original reactingFoam solver in the OpenFOAM ESI version assumes that species diffusivities are identical to thermal diffusivity. Further improvement by Cantera incorporation was demonstrated in \cite{li_quasi-direct_2020, yang_reactingfoam-sci_2019, zhou_robust_2022}. The reacting-flow mode in reactPlasFoam also benefits from the enhanced transport model provided by the Cantera library.

A 1D freely propagating stoichiometric hydrogen flame at atmospheric pressure and 300\,K was initially solved in Cantera using the mixture averaged transport model. The combustion chemistry is taken from Burke \emph{et al.} \cite{burke_comprehensive_2012}. The flame solution was subsequently utilized as the initial condition for reactPlasFoam.

\begin{figure}[H]
  \centering
  \includegraphics[width=0.95\linewidth]{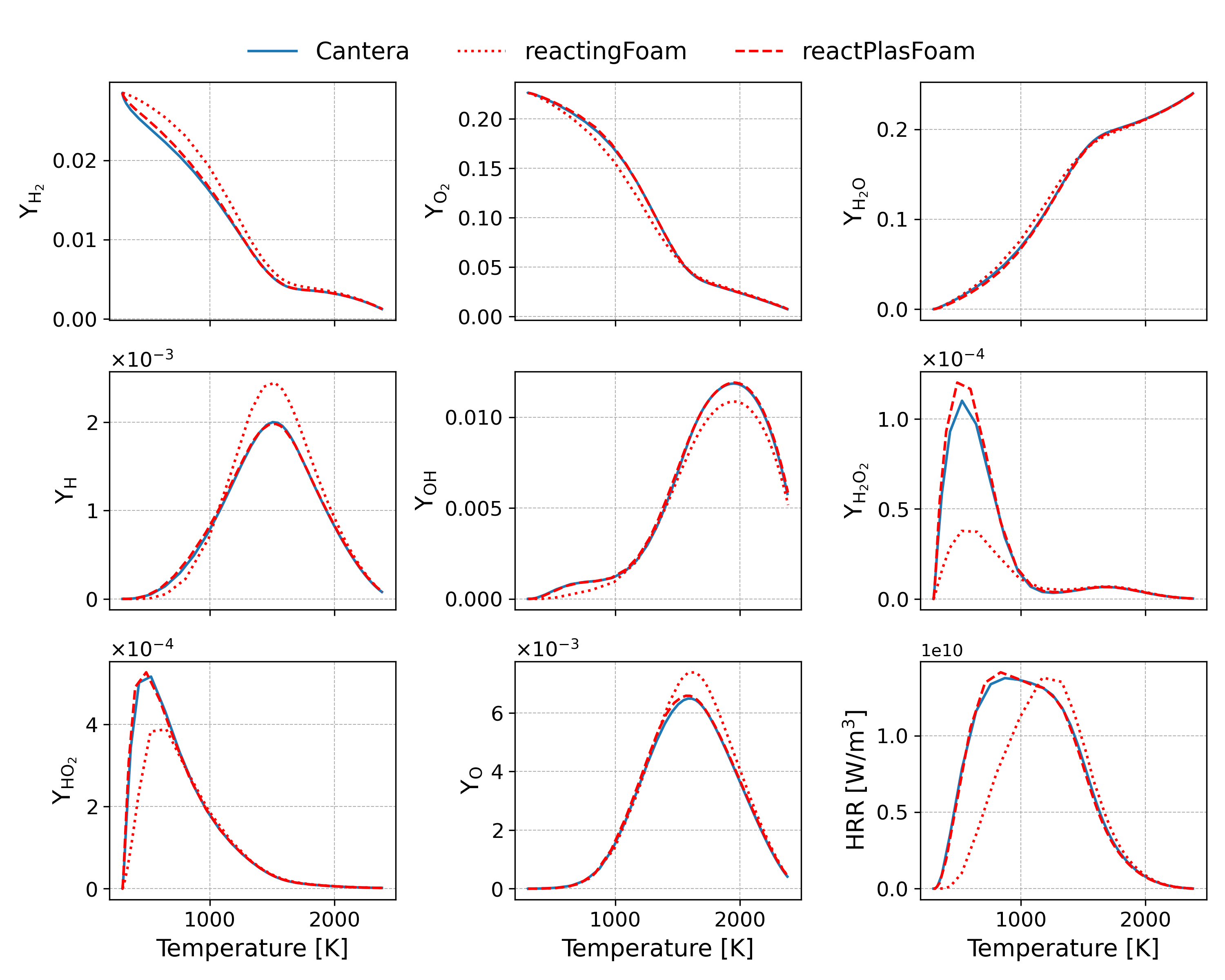}
  \caption{\label{fig:1Dflame} Comparison of solutions of 1D freely prorogation hydrogen flame in Cantera, reactingFoam and reactPlasFoam.}
\end{figure}

Figure~\ref{fig:1Dflame} illustrates the distributions of species mass fraction and volumetric heat release rate as functions of flame temperature. The original reactingFoam solver was also included in the comparison. A satisfying agreement between Cantera and reactPlasFoam is observed, while reactingFoam results in a significant deviation in radical distribution.

\newpage
\section{Applications and results} \label{applications}
This section reports various application examples of reactPlasFoam in realistic reacting flow conditions with detailed plasma and thermal chemical kinetics, demonstrating its capability in different operation modes: streamer, spark, flame, and ionic wind. The main goal is to highlight the unified model’s high performance, versatile capability, and significant potential to reveal intricate interactions between NTPs and reacting flows. Therefore, we do not delve into the detailed physical aspects in this paper. The secondary electrons emission on the electrodes is ignored. 

All three cases modeled in this section have experimental counterparts. The first case is a nanosecond spark discharge in preheated airflow, which constitutes the major part of code validation as it was well documented by Rusterholtz \emph{et al.}\ \cite{rusterholtz_ultrafast_2013}. The second case, a plasma-assisted methane flame, is inspired by the experimental setup in \cite{lacoste_analysis_2017}, demonstrating numerically for the first time streamer propagation inside a flame. The third case highlights the electrohydrodynamic force term in the model and exhibits flame displacement driven by an AC electric field \cite{park_dynamic_2018}.

A reduced GRI-3.0–based PAC mechanism developed for methane–air mixtures \cite{cheng_plasma_2022} is used in the first two cases. This mechanism has been well validated and contains 47 species participating in 429 elementary reactions. For the first case of air plasma, we retain 21 relevant species, which are marked in bold in Table \ref{tab:species}. The cross-section data for electron–neutral collisions are taken from the LXCat database \cite{pancheshnyi_lxcat_2012}, as detailed in Table \ref{tab:cs-data}.

\begin{table}[htbp]
  \centering
  \scriptsize            
  \setlength{\tabcolsep}{5pt}
  \renewcommand{\arraystretch}{1.5}
  \caption{Classification of species in the reduced methane–air PAC mechanism developed in \cite{cheng_plasma_2022} using GRI-3.0 as base combustion mechanism. Bold entries are retained in the air plasma simulations.}
  \label{tab:species}
  \begin{tabular}{@{} l p{0.75\textwidth} @{}}
  \toprule
  \textbf{Type} & \textbf{Species} \\
  \midrule
    Ground‑state neutrals &
      $\mathbf{N_{2}}$, $\mathbf{O_{2}}$, $\mathbf{O}$, $\mathbf{N}$,
      $\mathbf{NO}$, $\mathbf{NO_{2}}$, $\mathbf{O_{3}}$,
      $\mathrm{CH_{4}}$, $\mathrm{CO_{2}}$, $\mathrm{CO}$, $\mathrm{H_{2}O}$, $\mathrm{H}$, $\mathrm{OH}$, $\mathrm{H_{2}}$, $\mathrm{HO_{2}}$, $\mathrm{H_{2}O_{2}}$, 
      $\mathrm{CH}$, $\mathrm{CH_{2}}$, $\mathrm{CH_{3}}$, $\mathrm{C_{2}H_{4}}$, $\mathrm{C_{2}H_{5}}$, $\mathrm{C_{2}H_{6}}$, 
      $\mathrm{HCO}$, $\mathrm{CH_{2}O}$, $\mathrm{CH_{3}O}$, $\mathrm{CH_{3}OH}$, $\mathrm{H_{2}CN}$, $\mathrm{Ar}$ \\[2pt]
    
    Excited neutrals &
      $\mathbf{N_{2}(A)}$, $\mathbf{N_{2}(B)}$, $\mathbf{N_{2}(a)}$,
      $\mathbf{N_{2}(C)}$, $\mathbf{N(^{2}D)}$,
      $\mathbf{O_{2}(a^{1}\Delta_{g})}$, $\mathbf{O(^{1}D)}$,
      $\mathbf{O(^{1}S)}$, $\mathrm{CH_{2}(S)}$, $\mathrm{Ar^{*}}$ \\[2pt]
    
    Cations &
      $\mathbf{N_{2}^{+}}$, $\mathbf{O_{2}^{+}}$, $\mathbf{NO^{+}}$,
      $\mathrm{CH_{4}^{+}}$, $\mathrm{CH_{3}^{+}}$, $\mathrm{Ar^{+}}$ \\[2pt]
    
    Anions/electron &
      $\mathbf{O_{2}^{-}}$, $\mathbf{O^{-}}$, $\mathbf{e^{-}}$ \\
      \bottomrule
  \end{tabular}
\end{table}

\begin{table}[htbp]
    \centering
    \scriptsize            
    \setlength{\tabcolsep}{5pt}
    \renewcommand{\arraystretch}{1.5}
    \caption{Electron colliding partners considered in section \ref{airSpark} (bold entries) and section \ref{flameStreamer} (full entries). Cross section data is extracted from \cite{pancheshnyi_lxcat_2012}.}
    \label{tab:cs-data}
    \begin{tabular}{@{} p{0.45\textwidth} l @{}}
      \toprule
      \textbf{Neutral colliders} & \textbf{Data source} \\[2pt]
      \midrule    
       $\mathbf{N_2}$, $\mathbf{O_2}$, $\mathbf{O(^{3}P, \, ^{1}D)}$, $\mathbf{N(^{4}S, \, ^{2}D)}$, $\mathrm{CO_2}$, $\mathrm{CO}$, $\mathrm{H_2}$
       & IST-Lisbon \\[2pt]
       $\mathbf{N_2(A, B, C, a)}$, $\mathrm{Ar}$
       & Phelps \\[2pt]
       $\mathbf{O_2(a^{1}\Delta_{g})}$, $\mathrm{H_2O}$
       & Triniti  \\[2pt]
       $\mathbf{O_3}$
       & Morgan  \\[2pt]
       $\mathbf{NO}$, $\mathrm{CH_4}$
       & Hayashi  \\[2pt]
       
    \bottomrule    
    \end{tabular}

\end{table}

\subsection{Nanosecond spark discharge in air flow} \label{airSpark}
Fundamental studies of NRP discharges in air are essential given their wide applications in combustion, materials processing, and flow control. They also serve as critical validation for reactPlasFoam and similar codes to capture the ultrafast heating, radical production, and hydrodynamic response induced by spark discharges before tackling more complex reacting‐flow scenarios such as flames. To the authors’ knowledge, the only fully multiD simulation of spark discharges in air with detailed gas–plasma kinetics was conducted by Barleon \emph{et al.}\ using the AVIP–AVBP framework \cite{barleon_detailed_2022}. The challenges include the scarcity of suitable modeling tools and the problem’s inherent complexity—covering initial conditions, reaction mechanisms, electrode geometry, and power‐supply dynamics.

\begin{figure}[H]
  \centering
  \includegraphics[width=0.8\linewidth]{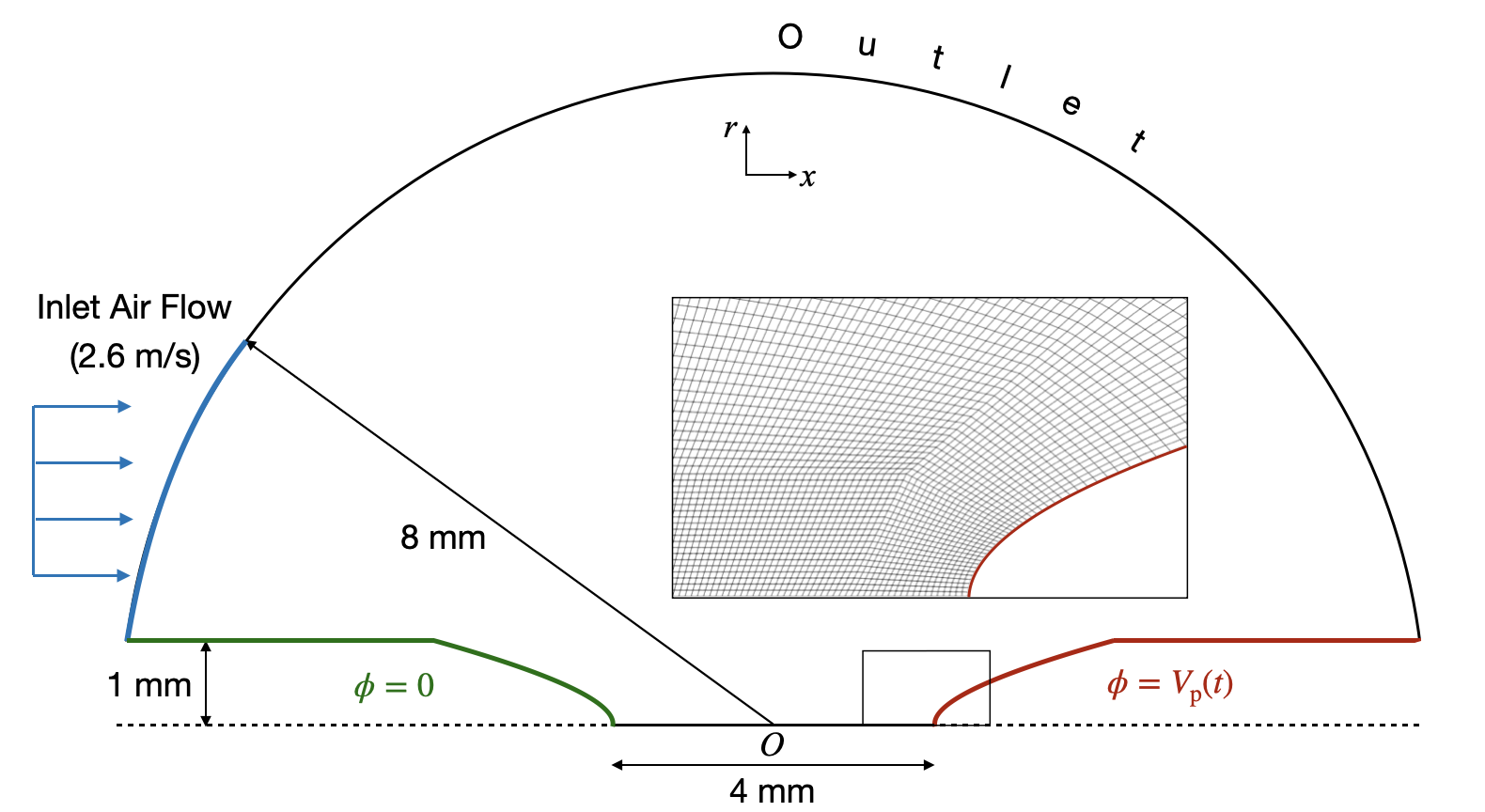}
  \caption{\label{fig:spark_config} Computational domain for nanosecond spark discharge in airflow. The baseline mesh near the anode has a minimum cell size of 25\,$\mu$m.}
\end{figure}

Our numerical setup is based on the experiments of Rusterholtz \emph{et al.}\ \cite{rusterholtz_ultrafast_2013}. Figure~\ref{fig:spark_config} shows the 2D axisymmetric domain: pin–pin electrodes separated by 4\,mm, immersed in a preheated ($1000\,$K) atmospheric‐pressure airflow. The electrodes have hyperbolic tips with a 200\,$\mu$m radius. The inlet flow, directed from cathode to anode at 2.6\,m/s, supplies fresh artificial air and removes plasma‐generated heat; its effect is minor here since only a single pulse is modeled. The applied voltage pulse $V_p$ peaks near 5.7\,kV, with waveform taken from \cite{rusterholtz_ultrafast_2013}. Zero‐gradient conditions are used for species and enthalpy at all walls, while $\nabla\phi=0$ is imposed on far‐field boundaries for Poisson’s equation. The baseline mesh near the anode (Figure~\ref{fig:spark_config}) has a minimum cell size of 25\,$\mu$m; AMR (Sec.~\ref{AMR}) refines this as needed. Note that the baseline mesh must resolve the hyperbolic curvature, as AMR simply splits cells by edge midpoints without geometric awareness.

Rusterholtz \emph{et al.}\ report quasi‐steady NRP spark conditions after multiple pulses, but modeling this transient requires hundreds of pulses and an external‐circuit model, posing a significant computational challenge \cite{barleon_detailed_2022}. Moreover, the published voltage waveform corresponds to the quasi‐steady state, whereas the first pulses of NRP discharges typically follow a different profile \cite{pancheshnyi_ignition_2006}. Consequently, we simulate only a single pulse under two initial‐condition sets: (1) a uniform $T=1000\,$K and a very low background ionization, representing the first pulse; (2) Gaussian profiles of temperature and pre‐ionization at the electrode gap, representing conditions at the onset of the next pulse in quasi‐steady operation.

\subsubsection{The very first pulse} \label{1stPulse}
\begin{figure}[H]
  \centering
  \includegraphics[width=0.95\linewidth]{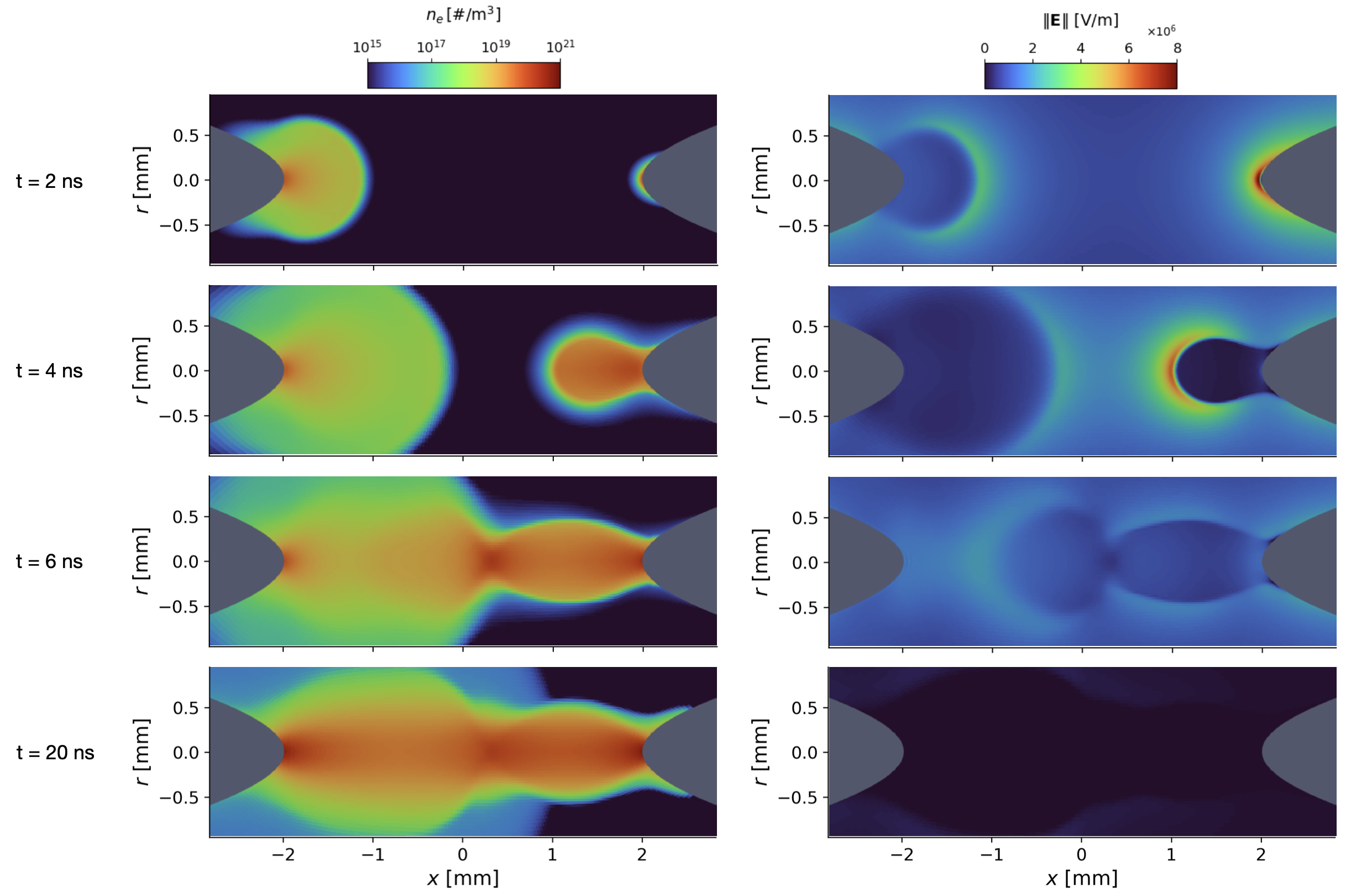}
  \caption{\label{fig:caseA_streamer} Streamer propagation in artificial air (N$_2$\,:\,O$_2$\;=\;0.79\,:\,0.21 in mole) with uniform initial conditions: $n_e=1\times10^9\,\mathrm{m^{-3}}$, $T=1000\,$K. Left column: electron number density; right column: electric field strength.}
\end{figure}

To reproduced the very first pulse behavior, a  low background electron density of $1\times10^9\,\mathrm{m^{-3}}$ is used, and the photoionization model is included. Figure~\ref{fig:caseA_streamer} shows streamer propagation via electron density and electric field strength at various times. At $t=2\,$ns, the negative streamer from the cathode has traveled 0.9\,mm, whereas the positive streamer has only just detached from the anode. This difference arises from their distinct propagation mechanisms: a positive streamer carries net positive space charge at its head and requires free electrons ahead of it, while a negative streamer propagates by drifting newly created electrons away from its head. Hence, low background ionization is unfavorable for positive‐streamer propagation. Moreover, Figure~\ref{fig:caseA_streamer} shows that the negative streamer is more diffuse and its head electric‐field strength is much weaker than that of the positive streamer, resulting in a lower ionization rate and smaller electron density within the negative streamer channel.

\begin{figure}[H]
  \centering
  \includegraphics[width=0.48\linewidth]{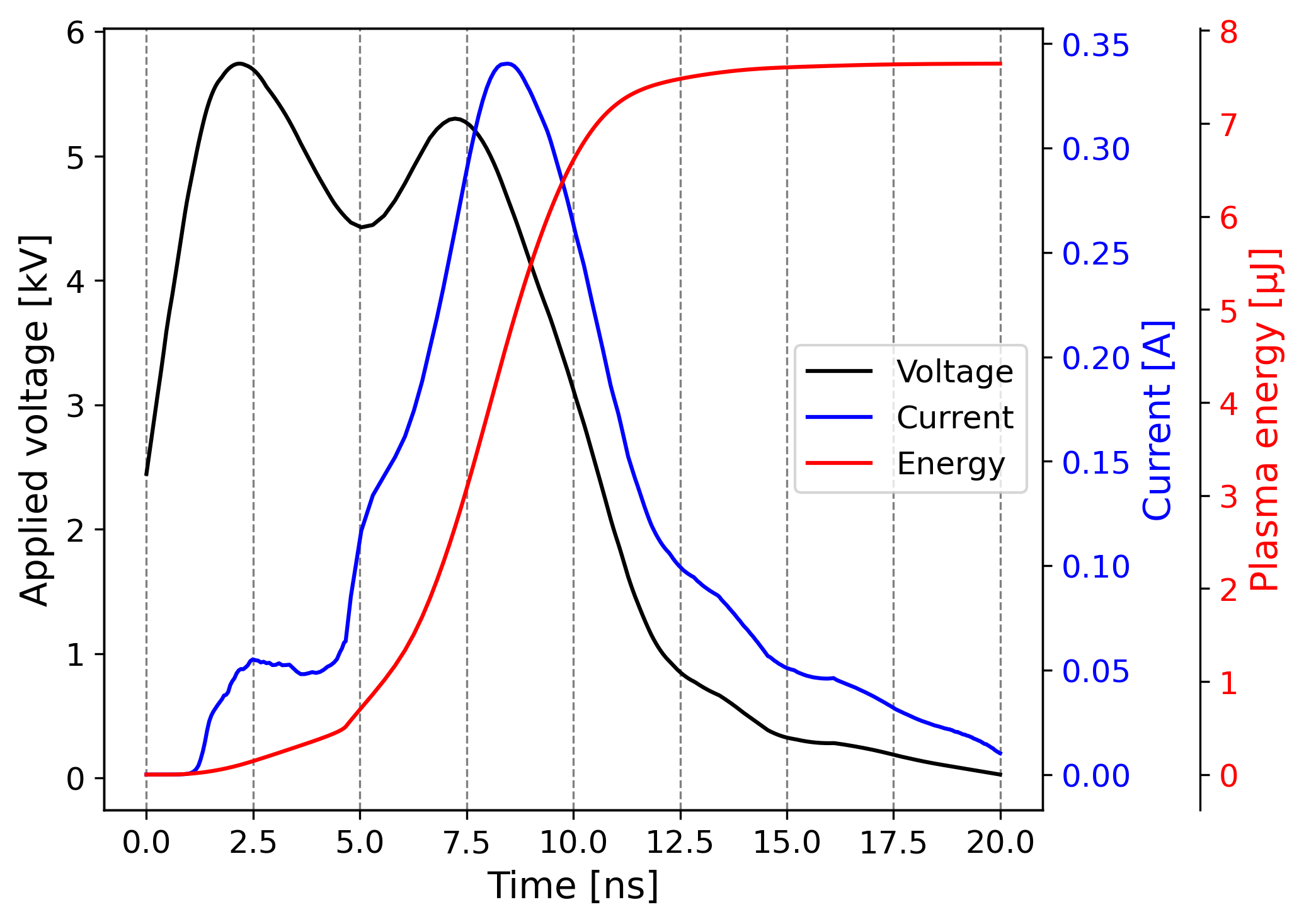}
  \caption{\label{fig:caseA_VI} Applied voltage, calculated conduction current, and deposited plasma energy for the very first pulse.}
\end{figure}

The two streamers connect and merge around $t=6\,$ns, forming a conductive channel and causing the conduction current to surge (Figure~\ref{fig:caseA_VI}). The conduction current is calculated over a cross‐section $S_f$ spanning the electrode gap near the anode:
\begin{equation}
  I_{\rm cond} = \iint_{S_f} e\,n_e\,\mu_e\,\mathbf{E}\cdot d\mathbf{A}.
  \label{eq:cond_current}
\end{equation}

This glow discharge does not transition into a spark discharge, since the applied voltage begins to fall at $t=7.5\,$ns and the current cannot rise further. The total energy deposited is only about $7\,\mu$J, far less than the $670\,\mu$J measured in the spark regime by Rusterholtz \emph{et al.}\ \cite{rusterholtz_ultrafast_2013}. Note that the voltage waveform here matches the quasi‐steady state, whereas the first few NRP pulses often exhibit higher peak voltages than subsequent pulses \cite{pancheshnyi_ignition_2006}.

\begin{figure}[H]
  \centering
  \includegraphics[width=0.8\linewidth]{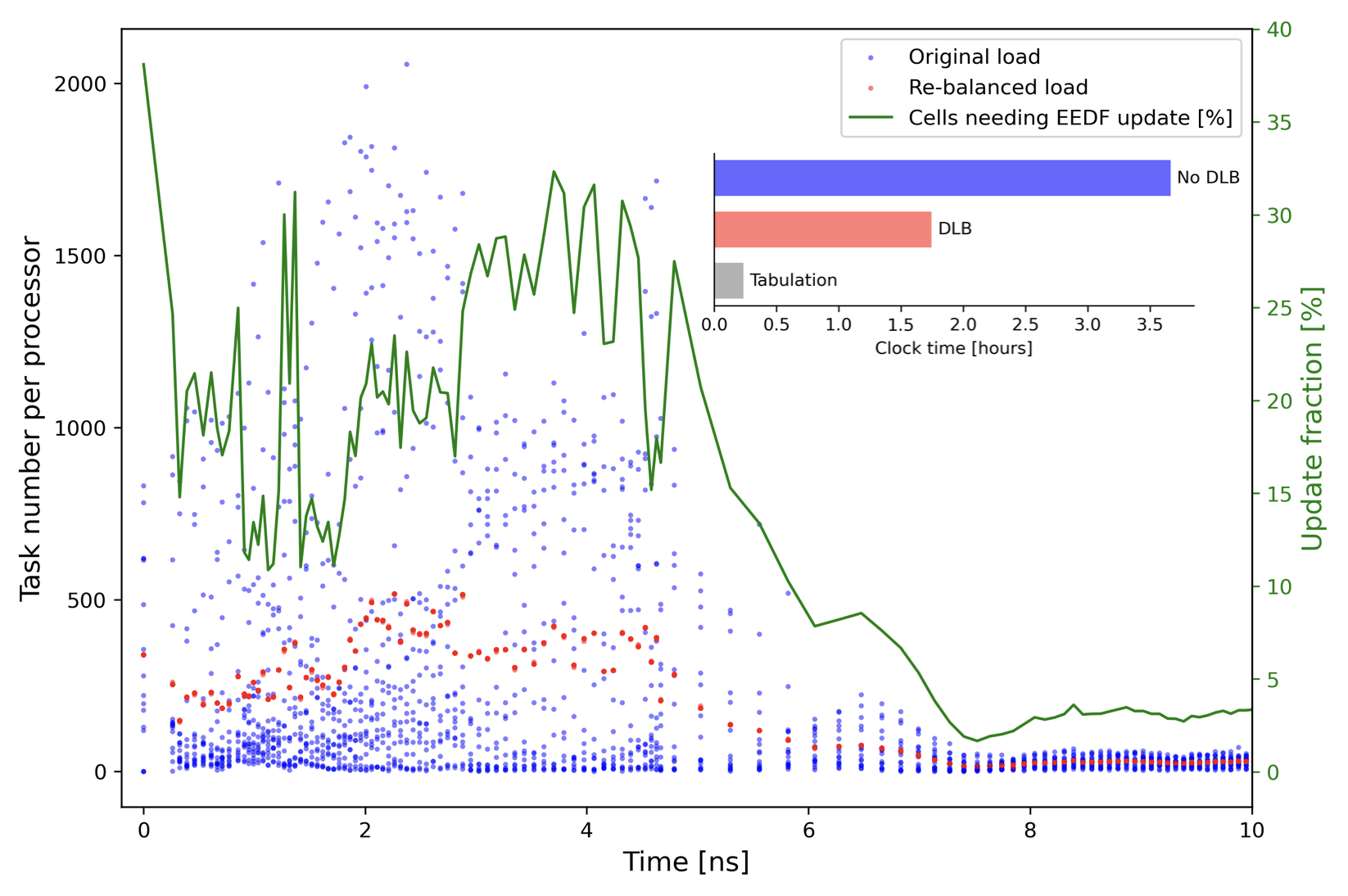}
  \caption{\label{fig:caseA_DLB} Distribution of EBE‐solve tasks per core on 16 cores (dots) sampled every 100 time steps, and total run time to 10 ns for on‐the‐fly vs.\ tabulated EBE.}
\end{figure}

The performance of on-the-fly EBE solves with dynamic load balancing is illustrated in Figure~\ref{fig:caseA_DLB}. Each dot represents the number of EBE tasks (i.e., the subdomain size of $\mathcal{C}$) assigned to one of 16 cores at a given time. The wide scatter of the original (blue) dots indicates severe load imbalance, arising from the localized evolution of the electric field during streamer propagation in different subdomains. For example, a core covering a far‐field region experiences fewer EBE updates and becomes underloaded.

With dynamic load balancing (red dots), tasks are redistributed each time step, yielding a much narrower bandwidth that nearly collapses onto a single value. The initial imbalance diminishes after the positive and negative streamers connect, since the electric‐field changes then span the entire discharge channel rather than isolated pockets.

Applying DLB cuts the EBE‐solve cost roughly in half. The overall speedup is modest, however, because the update fraction (the percentage of cells in $\mathcal{C}$) remains high. When nearly all cells require EBE solves, there is little imbalance to correct. Indeed, if one solves EBE on every cell, DLB becomes unnecessary, as AMR has already balanced cell counts across cores. The computational domain and meshing strategy in this air discharge simulation have been optimized, allocating very limited grid number for the region without plasma. In more extreme multi‐scale problems where plasma occupies only a small portion of a large domain, the update fraction would be lower and DLB more effective; this is demonstrated in Sec.~\ref{flameStreamer}.

Since the mixture composition is homogeneous and changes little during the first pulse, a lookup table for electron properties is also suitable. Figure~\ref{fig:caseA_DLB} shows that using tabulation instead of on-the-fly EBE lowers the 10\,ns run time to about 15\,min, with negligible loss in accuracy. This underscores the significant cost of online EBE solutions.

\subsubsection{Effects of initial \texorpdfstring{$T$}{T} and \texorpdfstring{$n_e$}{ne} profiles}
\begin{figure}[H]
  \centering
  \includegraphics[width=0.48\linewidth]{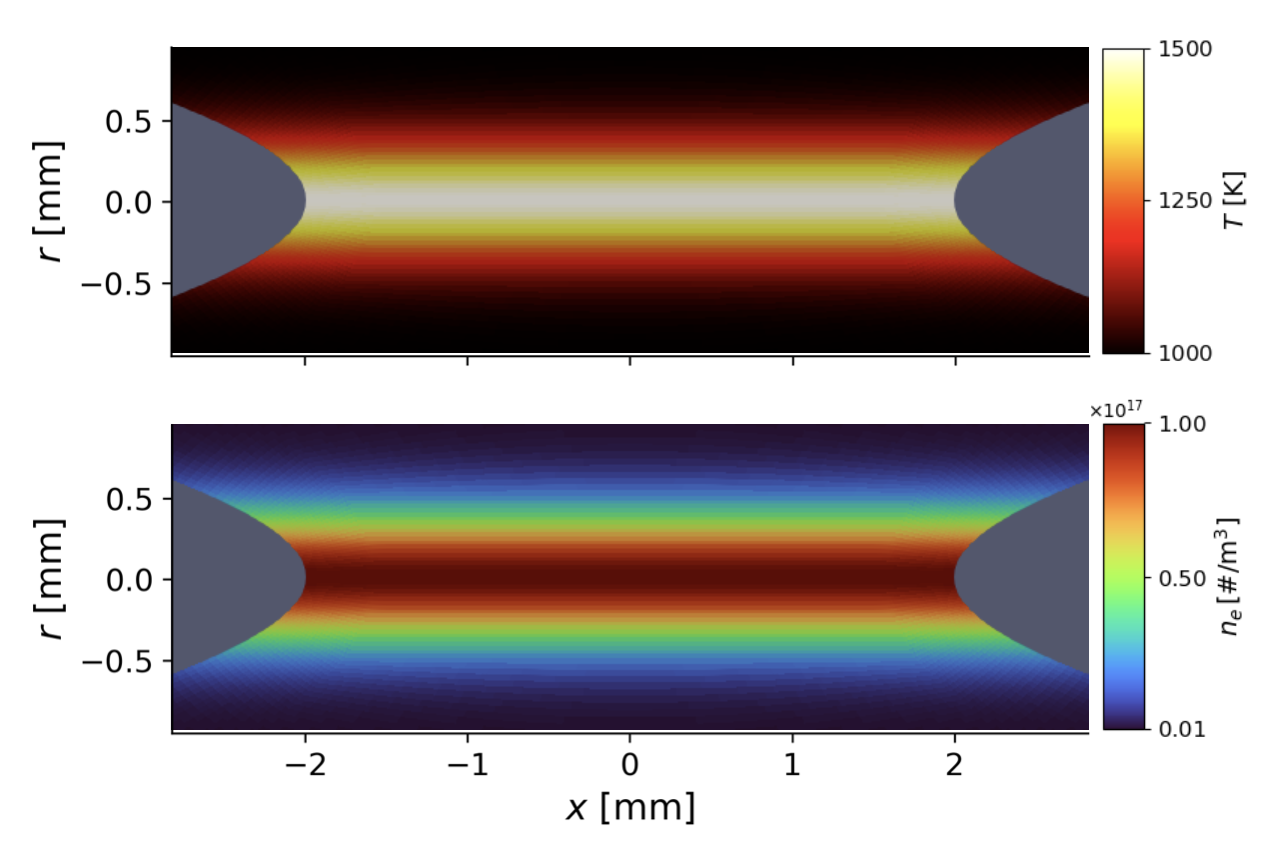}
  \caption{\label{fig:caseD_initial} Initial radial profiles of temperature and electron number density accounting for previous pulses.}
\end{figure}

Experimental measurements in \cite{rusterholtz_ultrafast_2013} show that, during the quasi‐steady state before the next voltage pulse, the gas temperature and atomic‐oxygen (O($^3$P)) number density at the electrode‐gap center are 1500\,K and $2\times10^{17}\,\mathrm{cm^{-3}}$, respectively. Previous discharges also leave a high background ionization, especially in the central channel region. Following Barleon \emph{et al.}\ \cite{barleon_detailed_2022}, we impose radial Gaussian profiles for temperature and electron density in the gap:
\begin{align}
n_e(x,r) &= 1\times10^{15} + 1\times10^{17}\exp\!\bigl(-r^2/\sigma_r^2\bigr)\,,\\
T(x,r)   &= 1000 + 500\exp\!\bigl(-r^2/\sigma_r^2\bigr)\,,
\end{align}
with $\sigma_r=400\,\mu\mathrm{m}$ to account for interpulse diffusion. Atomic‐oxygen also follows a Gaussian distribution, yielding a gap‐center mole fraction N$_2$:O$_2$:O = 0.774:0.186:0.04. Figure~\ref{fig:caseD_initial} shows these initial profiles.

\begin{figure}[H]
  \centering
  \includegraphics[width=0.95\linewidth]{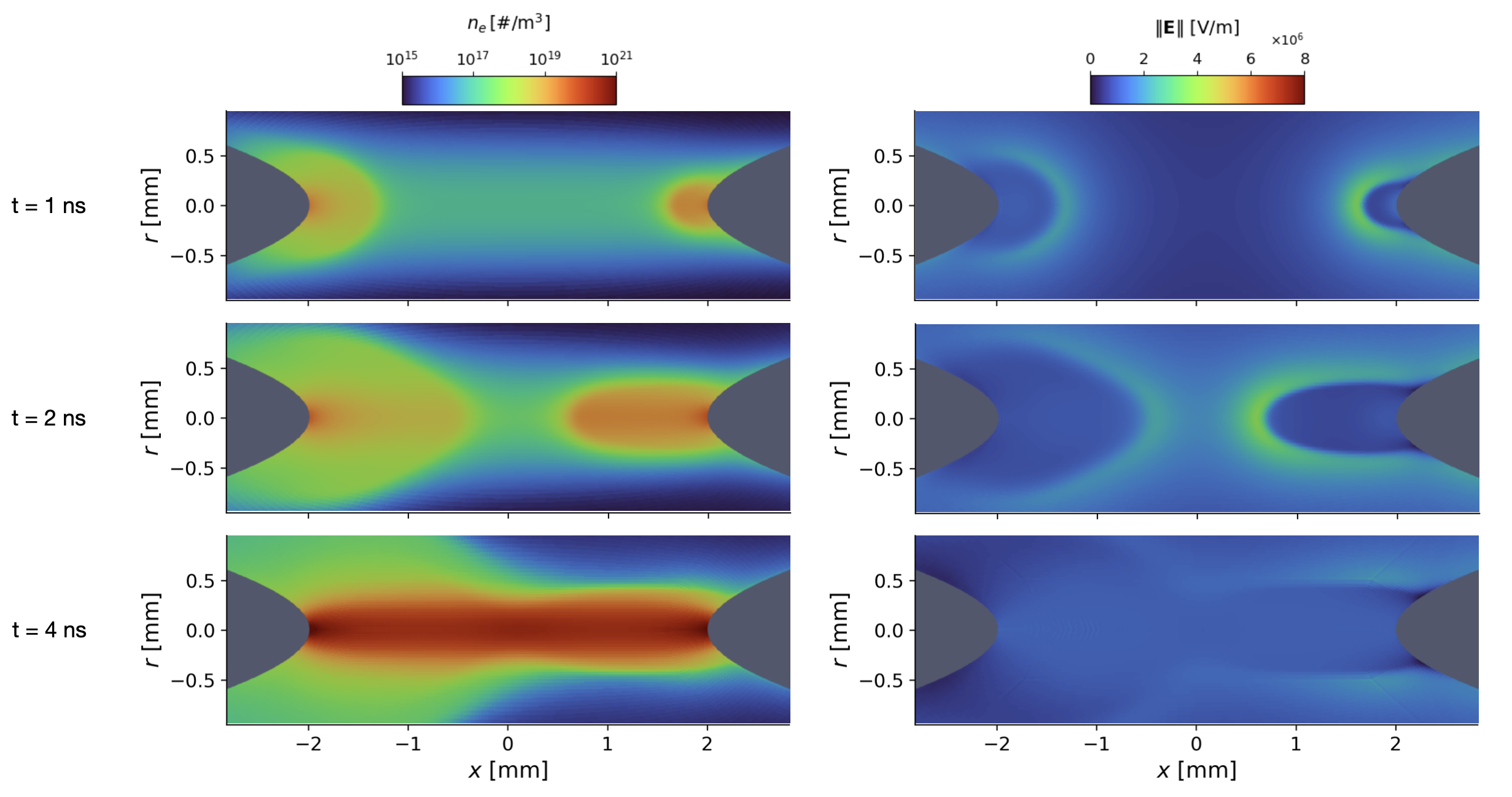}
  \caption{\label{fig:caseD_streamer} Streamer propagation in air with high pre‐ionization and background heating.}
\end{figure}

\begin{figure}[H]
  \centering
  \includegraphics[width=0.48\linewidth]{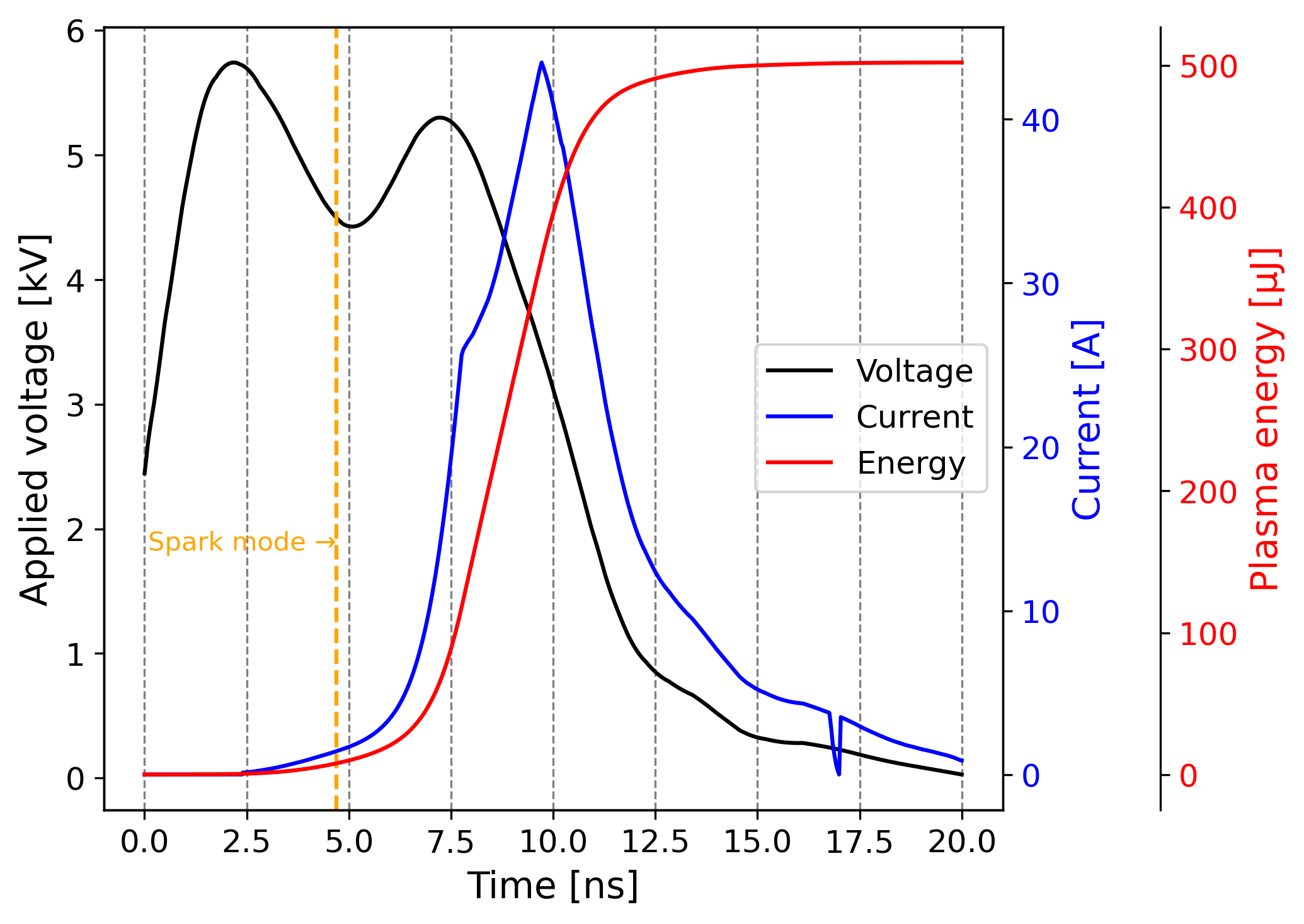}
  \caption{\label{fig:caseD_VI} Applied voltage, conduction current, and deposited plasma energy of the spark discharge. Spark mode is triggered at $t=4.69\,$ns (orange dashed line).}
\end{figure}

Figure~\ref{fig:caseD_streamer} presents streamer propagation under these conditions, showing both negative and positive streamers traveling faster than in the very first pulse. Elevated temperature increases the reduced field near the axis, and high pre‐ionization facilitates positive‐streamer propagation and smooths the streamer heads. A narrow conductive channel forms as early as $t=4\,$ns, transitioning into spark mode at $t=4.9\,$ns. Figure~\ref{fig:caseD_VI} plots the conduction current and deposited plasma energy. Note that the external‐power cap (Eq.~\ref{eq:spark_ne}) prevents excessive electron densities. The total deposited energy reaches 502\,µJ at 24\,ns, about 25\% below the experimental $670\pm50\,$µJ \cite{rusterholtz_ultrafast_2013}. The energy fractions corresponding to heating and vibration are 20\% and 27\%, respectively. Using tabulation for the EBE solution, the 25\,ns simulation completes within seven minutes on 16 cores, thanks to the reduced streamer‐head stiffness and early spark transition.

\begin{figure}[H]
  \centering
  \includegraphics[width=0.48\linewidth]{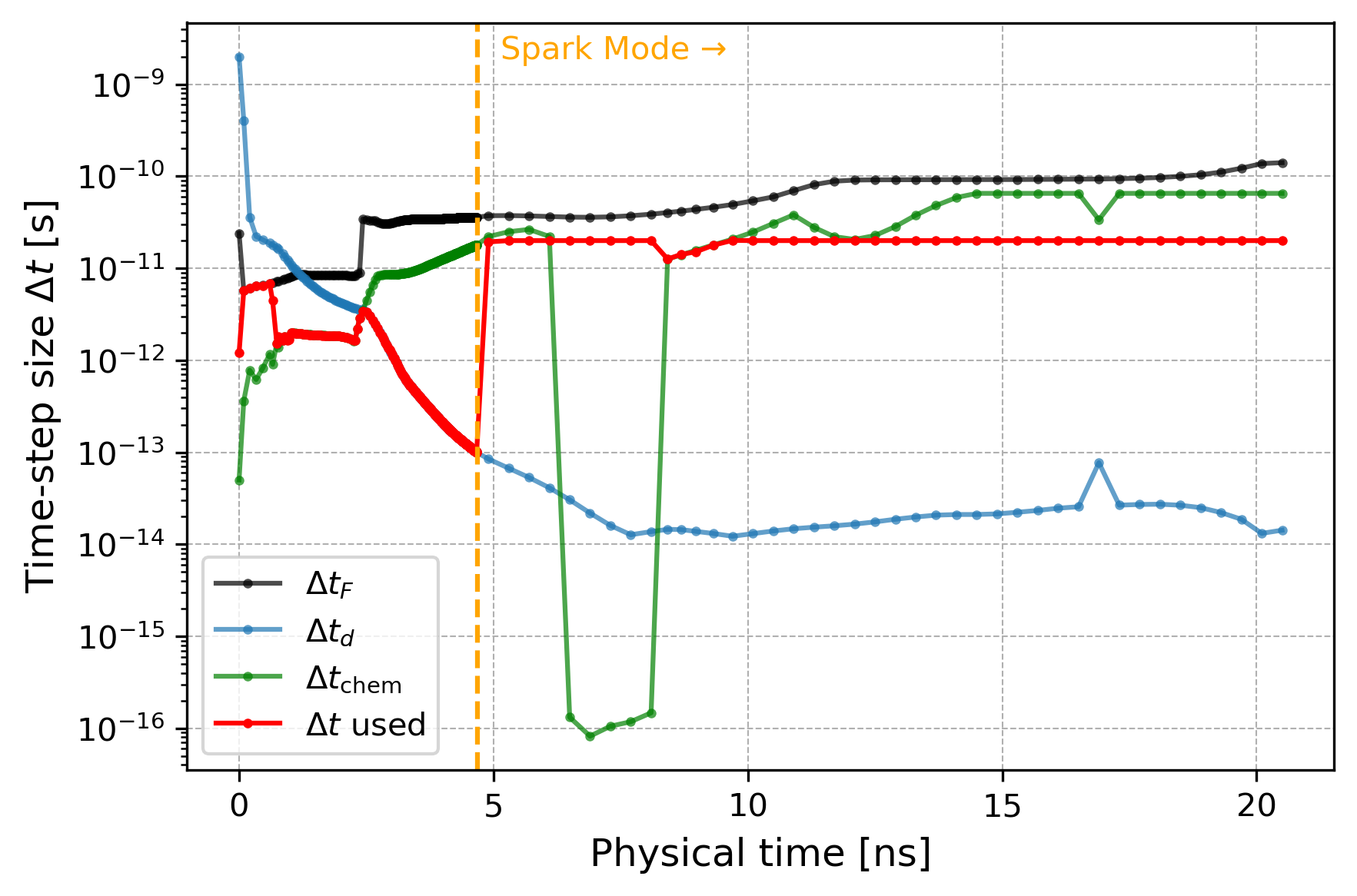}
  \caption{\label{fig:caseD_dt} Time‐step constraints versus physical time for spark discharge modeling (samples every 20 time steps).}
\end{figure}

The time‐step constraints in the spark discharge simulation are shown in Figure~\ref{fig:caseD_dt}. Initially, when $t<0.7\,$ns, the time step $\Delta t$ is limited by the Fourier condition. As the discharge develops, the chemical time scale becomes the most restrictive factor. At $t=2.5\,$ns a connected channel forms (see the conduction‐current rise in Figure~\ref{fig:caseD_VI}), causing $\Delta t_d$ to drop from approximately $10^{-12}\,$s to $10^{-14}\,$s. Spark mode is triggered when $\Delta t = \Delta t_d = 1\times10^{-13}\,$s, improving efficiency at the cost of exact electron conservation. During spark mode, the time step remains fixed at $2\times10^{-11}\,$s even though $\Delta t_d$ is smaller. Note that a sudden drop in $\Delta t_{\rm chem}$ may reflect numerical artifacts in the implementation of Eq.~(\ref{dt_c}).

\begin{figure}[H]
  \centering
  \includegraphics[width=0.95\linewidth]{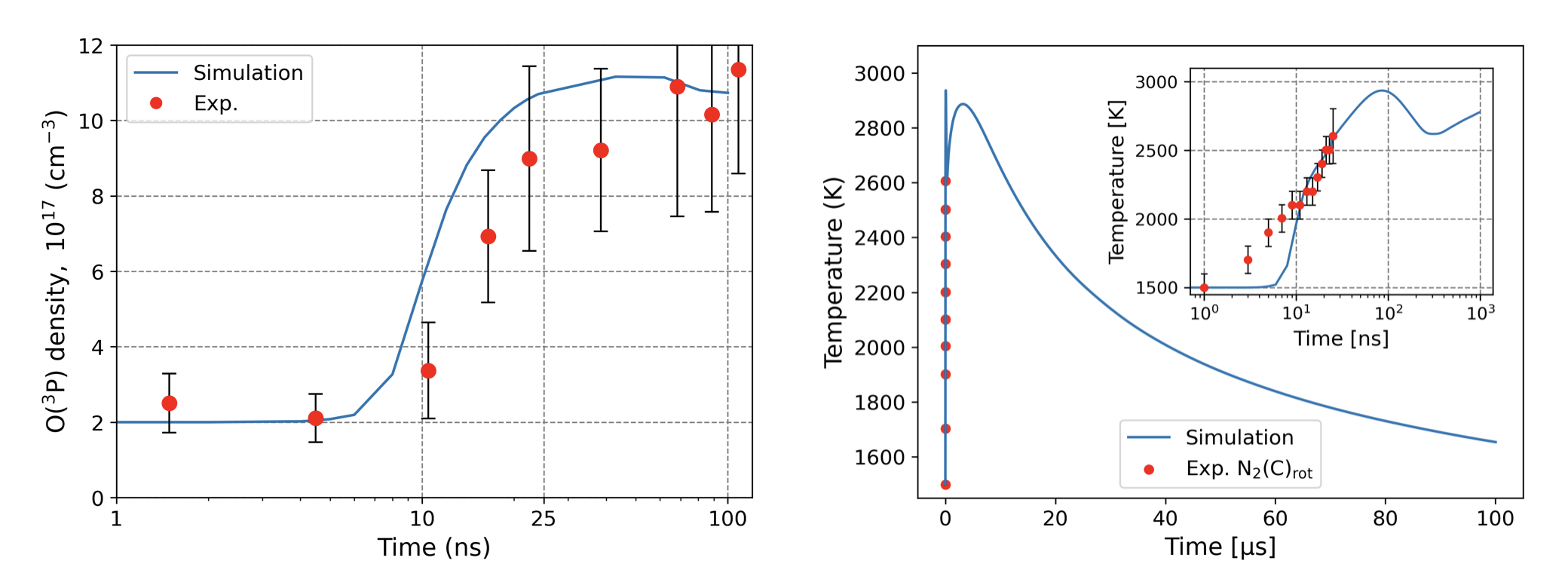}
  \caption{\label{fig:caseD_T_nO} Temporal evolution at the electrode‐gap center of temperature and O($^3$P) density: simulation vs.\ experiment \cite{rusterholtz_ultrafast_2013}.}
\end{figure}

\begin{figure}[H]
  \centering
  \includegraphics[width=0.48\linewidth]{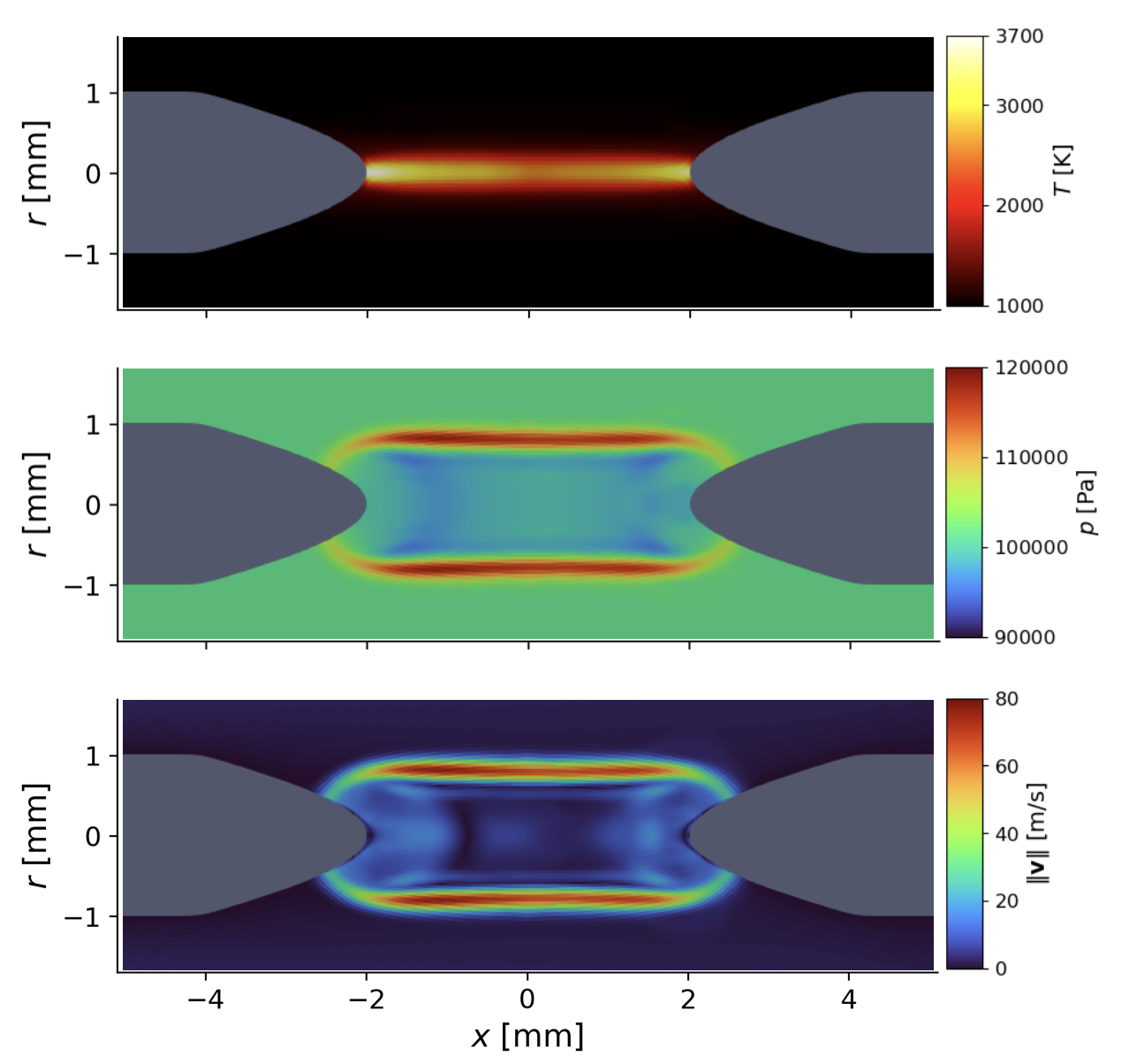}
  \caption{\label{fig:caseD_wave} 2D maps of temperature, pressure, and velocity magnitude at $t=1\,\mu$s.}
\end{figure}

Figure~\ref{fig:caseD_T_nO} compares simulated gas temperature and O($^3$P) density at the discharge center $(x,r)=(0,0)$ with the experimental data from \cite{rusterholtz_ultrafast_2013}. The agreement validates reactPlasFoam’s ability to capture rapid gas heating and radical production in multiD simulations. The temperature profile shows a sharp drop of about 300\,K after $t=100\,$ns, due to plasma‐kernel expansion. Figure~\ref{fig:caseD_wave} presents the pressure‐wave expansion at $t=1\,\mu$s. Temperature is highest near the electrode tips because electron and energy densities peak there. This nonuniformity diffuses but remains evident at $1\,\mu$s. Thereafter, the kernel temperature evolves under thermal diffusion and V–T relaxation, rising until $t=5\,\mu$s and then decreasing slowly until the end of the pulse interval.

\newpage
\subsection{Streamer propagation in a flame} \label{flameStreamer}
\begin{figure}[H]
\centering
\includegraphics[width=0.9\linewidth]{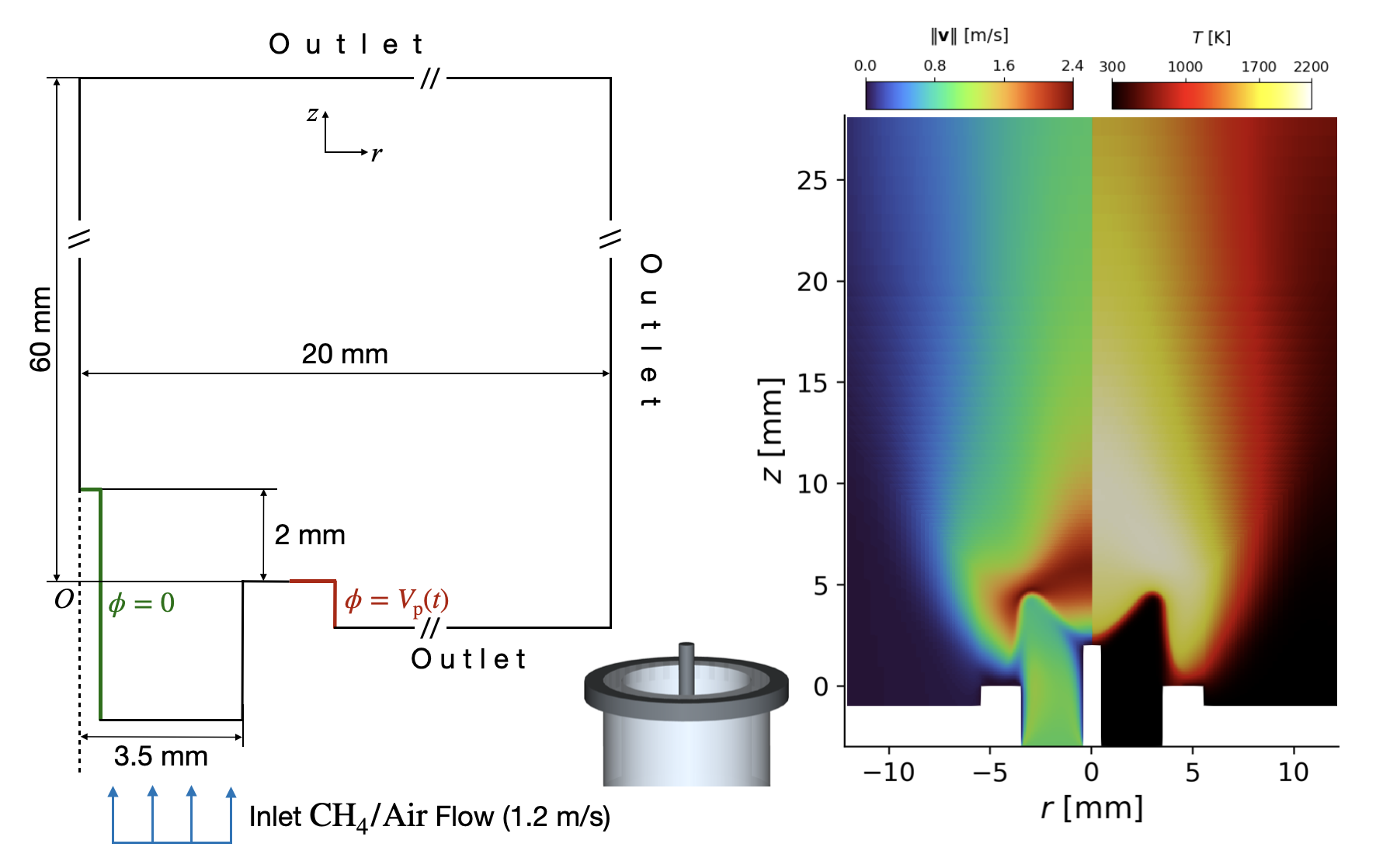}
\caption{\label{fig:rodring-config} Schematics of a laminar premixed methane flame with rod–ring electrodes. Left: 2D axisymmetric computational domain. Right: steady‐state velocity magnitude and temperature contours of the base flame.}
\end{figure}

In the experiments of Lacoste \emph{et al.} \cite{lacoste_analysis_2017}, a jet burner with a rod–ring electrode configuration was used to analyze the step response of a laminar premixed methane flame to plasma actuation. The corresponding numerical setup is shown in Figure~\ref{fig:rodring-config}. The quartz burner tube is 1\,mm thick and has a 3.5\,mm inner diameter. A grounded rod electrode of diameter 0.85\,mm protrudes 2\,mm beyond the burner lip. A ring electrode surrounds the burner’s upper surface and receives the high-voltage pulse. The 2D axisymmetric domain is supplied with premixed CH$_4$/air at 1.2\,m/s and equivalence ratio 0.95. The simulated base flame, which stabilizes into an “M” shape and closely matches the experiments in \cite{lacoste_analysis_2017}, serves as the initial condition for streamer propagation.

The PAC mechanism from \cite{cheng_plasma_2022} is better suited to ignition problems than to fully detailed flames, as it omits plasma kinetics in the reaction zone and burned gas. Incorporating and validating the complete mechanism is beyond this study’s scope. To compensate, we add four collision partners, CO$_2$, CO, H$_2$, and H$_2$O, to the EEDF cross‐section database (Table~\ref{tab:cs-data}) as compared the original list in \cite{cheng_plasma_2022}. Because the flame mixture is heterogeneous, the EBE must be solved simultaneously.

In reality, the reaction zone of a flame is extremely weakly ionized by itself. However, the current PAC mechanism does not include ion chemistry of combustion. Instead, we use a uniform low background electron density of $10^9\,\mathrm{m^{-3}}$ as the initial condition. A 8 kV trapezoid pulse is applied at the ring electrode including 2\,ns rise,  8\,ns plateau and 2\,ns fall.

\begin{figure}[H]
\centering
\includegraphics[width=0.95\linewidth]{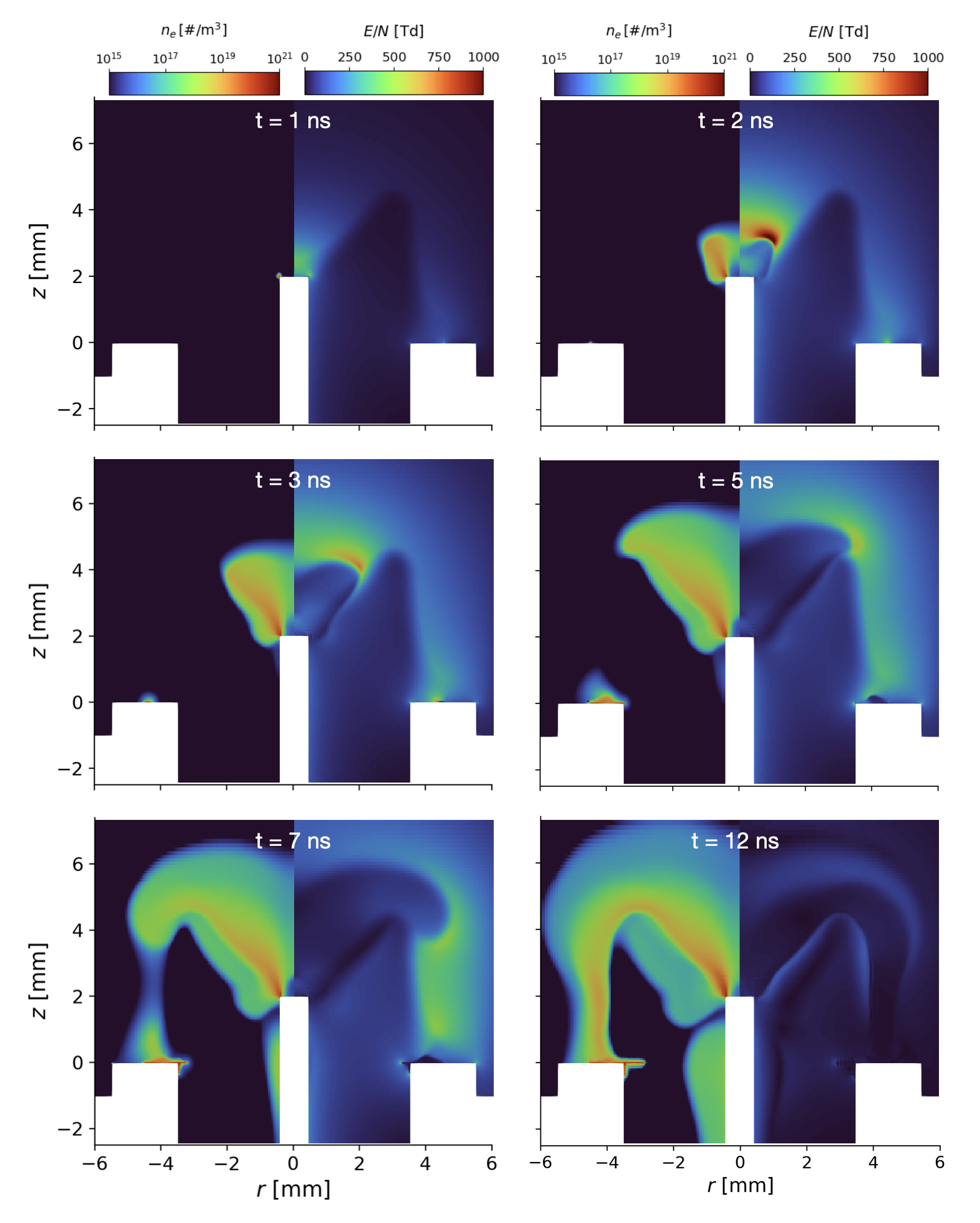}
\caption{\label{fig:rodring-streamer} Streamer propagation guided by the premixed methane/air flame front. Applied voltage: 8 kV.}
\end{figure}

The temporal evolution of the discharge is shown in Figure~\ref{fig:rodring-streamer}. Due to the presence of the flame, the spatial distribution of the reduced electric field is strongly modified. At $t=1\,$ns, the $E/N$ map clearly follows the flame curvature, indicating a jump in temperature (thus the inverse of gas number density) across the flame front. The negative streamer propagates from the central rod, following the enhanced reduced‐field region at the flame rather than the shortest path to the anode. The positive streamer appears only at $t=3\,$ns due to the lack of background electrons, consistent with Sec.~\ref{1stPulse}. A secondary discharge path emerges around $t=7\,$ns when the positive streamer reaches the burner lip and moves horizontally toward the cathode. Simultaneously, a weak, highly diffusive negative streamer develops on the lateral surface of the cathode. Before this secondary discharge fully develops, the applied voltage is cut off.

The major limitation of this 2D axisymmetric simulation is that experimental discharges between rod–ring electrodes tend to be filamentary and not perfectly axisymmetric. Streamer propagation is highly nonlinear and extremely sensitive to initial conditions and the background fields. For example, stochastic photoionization can trigger branching \cite{wang_quantitative_2023}, and Laplacian instability may arise deterministically \cite{arrayas_spontaneous_2002}. The validity of the 2D simplification  needs further examination in future work.

\begin{figure}[H]
\centering 
\includegraphics[width=0.8\linewidth]{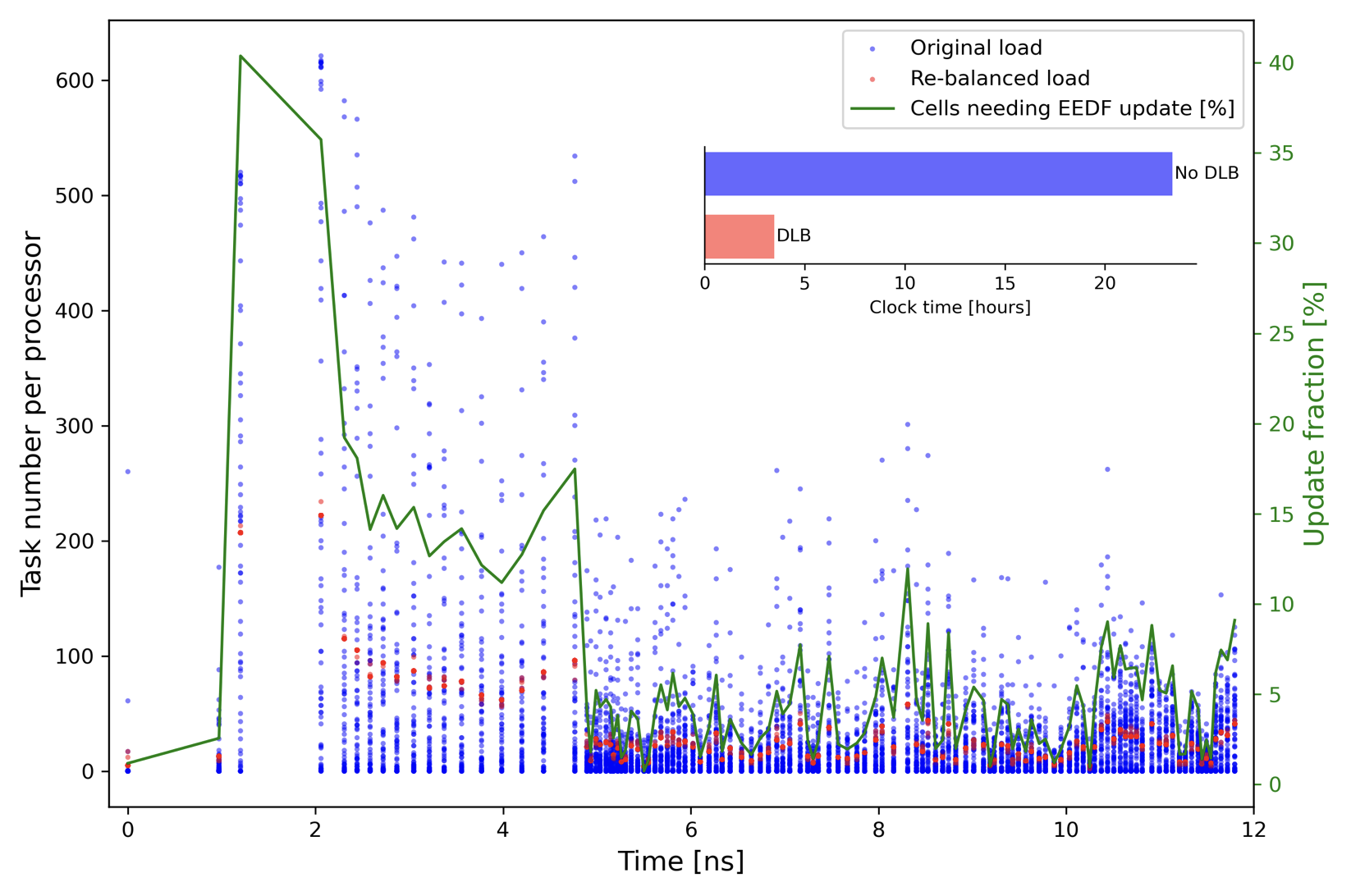}
\caption{\label{fig:rodring-DLB} Distribution of EBE‐solve tasks per core on 64 cores (dots) sampled every 100 time steps, and total run time to 12\,ns with/without DLB.}
\end{figure}

Since this study focuses on model and solver development rather than detailed physics, deeper exploration of two-way coupling between plasma and flame is left for future studies. This numerical experiment demonstrates reactPlasFoam's capability to handle complex NTP–reacting‐flow systems at reasonable cost. The efficiency of DLB for on‐the‐fly EBE solves is illustrated in Figure~\ref{fig:rodring-DLB}. The CPU‐hour consumption for the 12\,ns simulation with DLB is 222\,CPU\,h (3.47\,h × 64 cores), a sevenfold speedup compared to the run without DLB. This speedup exceeds that in Sec.~\ref{1stPulse}, owing to the lower EEDF update fraction and greater initial load imbalance (Fig.~\ref{fig:caseA_DLB}). Note that cost is higher under this low pre‐ionization condition and is expected to be lower for subsequent NRP discharges.

\newpage
\subsection{Flame under electric field} \label{efieldFlame}
Flames are weakly ionized plasmas, with ions and electrons formed via chemi-ionization reactions \cite{bisetti_calculation_2012}. The effect of a non‐breakdown external electric field on flames has been studied experimentally and numerically \cite{park_bidirectional_2016, park_dynamic_2018, belhi_modelling_2013, belhi_computational_2017}. Here, we examine the performance of the time‐integration sub‐cycling technique by applying the ionic-wind mode (Sec.~\ref{ionic_mode}) to model the dynamic response of a counterflow flame to an AC electric field.

\begin{figure}[H]
\centering
\includegraphics[width=0.8\linewidth]{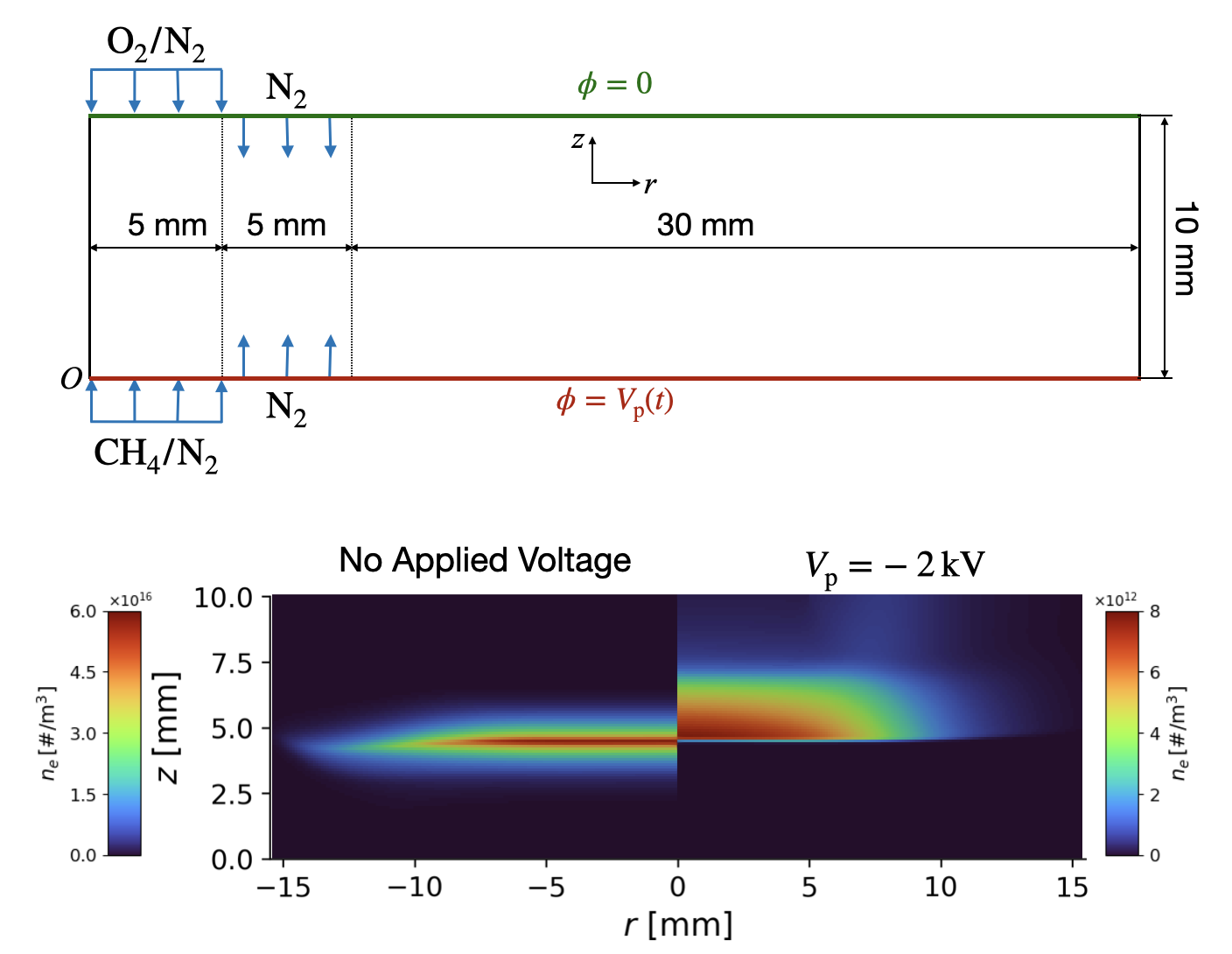}
\caption{\label{fig:Eflame-config} Computational domain reproduced from \cite{belhi_computational_2017} and 2D electron‐density maps with and without an applied DC voltage.}
\end{figure}

The computational domain and numerical setup are identical to those in \cite{belhi_computational_2017} and are reproduced in Figure~\ref{fig:Eflame-config}. The 2D axisymmetric domain represents the counterflow methane/air flame used in \cite{park_bidirectional_2016}. Opposed nozzles are separated by 1\,cm: nitrogen‐diluted methane enters from the bottom, oxygen from the top, and nitrogen co‐flows through outer nozzles of 2\,cm diameter. All inlets have a velocity of 20\,cm/s and a temperature of 300\,K at one atmosphere. The skeletal mechanism comprises 25 species, including electrons and six ions (H$_3$O$^+$, O$_2^-$, O$^-$, OH$^-$, CO$_3^-$, CHO$_3^-$). Figure~\ref{fig:Eflame-config} shows electron distributions with and without a DC field. With no applied voltage, electrons reside in the thin reaction zone layer, peaking at $\sim6\times10^{16}\,\mathrm{m^{-3}}$. A DC field drives electrons toward the top nozzle, spreading them into the N$_2$/O$_2$ flow and reducing the peak density to below $1\times10^{13}\,\mathrm{m^{-3}}$.

\begin{figure}[H]
\centering
\includegraphics[width=0.48\linewidth]{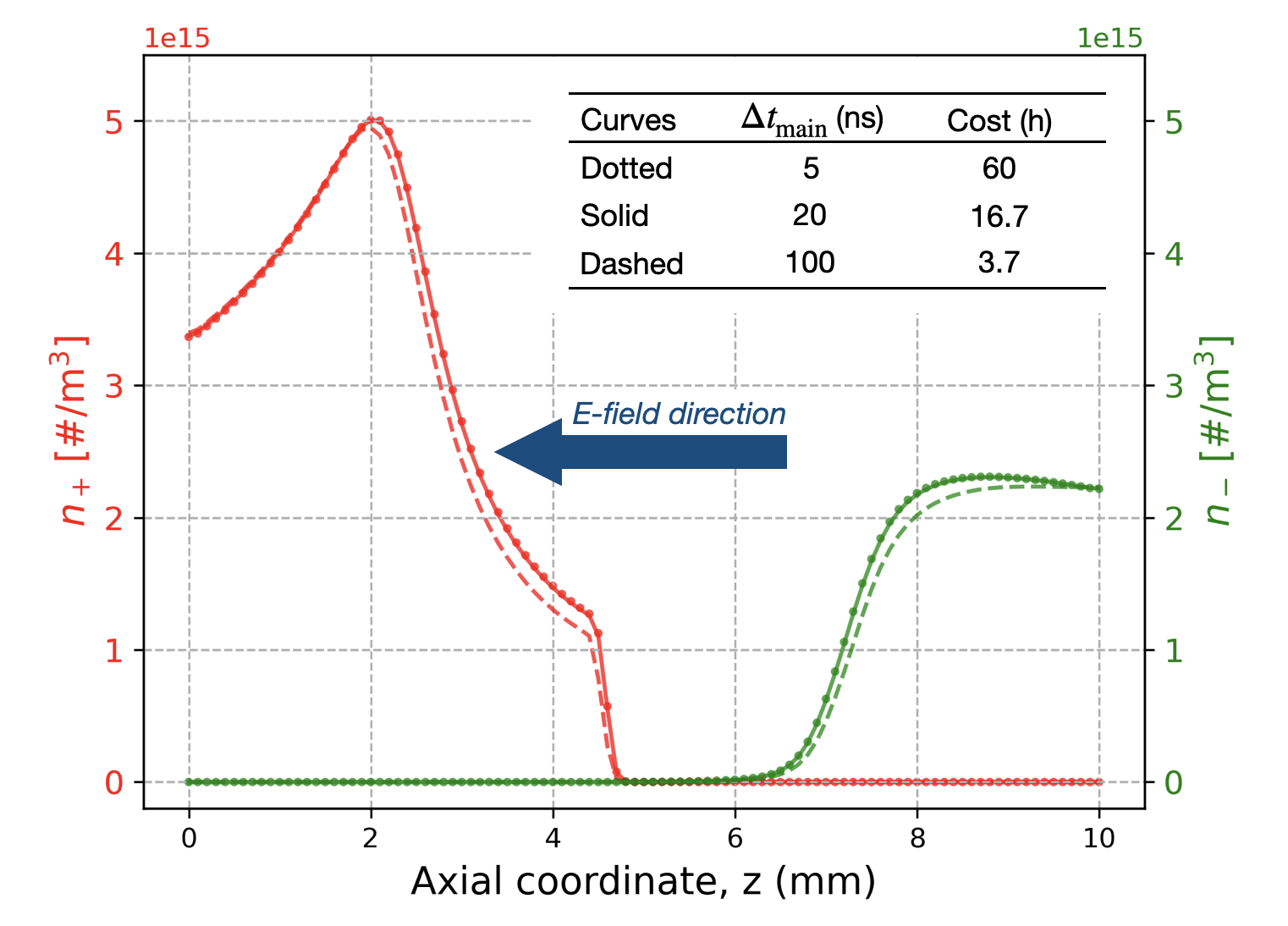}
\caption{\label{fig:Eflame-subcycle} Ion densities along the axis under $V_\text{p}=-2\,$kV. Sub‐timestep $\Delta t_\text{sub}=5\,$ns; main timestep $\Delta t_\text{main}$ varied. All runs on 256 cores.}
\end{figure}

Since we have direct access to the code and case files of \cite{belhi_computational_2017}, reproducing their simulation results would be redundant. Instead, we focus on validating the subcycling speedup. Figure~\ref{fig:Eflame-subcycle} plots ion densities along the central axis under a DC field of $V_\text{p}=-2\,$kV. Dotted curves are the benchmark using a uniform $\Delta t=5\,$ns for all equations, which requires 60\,h on 256 cores to simulate 5\,ms of physical time. By fixing the sub‐timestep for charged‐species transport and Poisson’s equation at $\Delta t_\text{sub}=5\,$ns and increasing outer loop time step, significant speedup can be achieved with acceptable loss of accuracy. We found that the computational cost can be reduced by 16 folds with $\Delta t_\mathrm{main}$ relaxed to 0.1 ms.

\begin{figure}[H]
\centering
\includegraphics[width=0.48\linewidth]{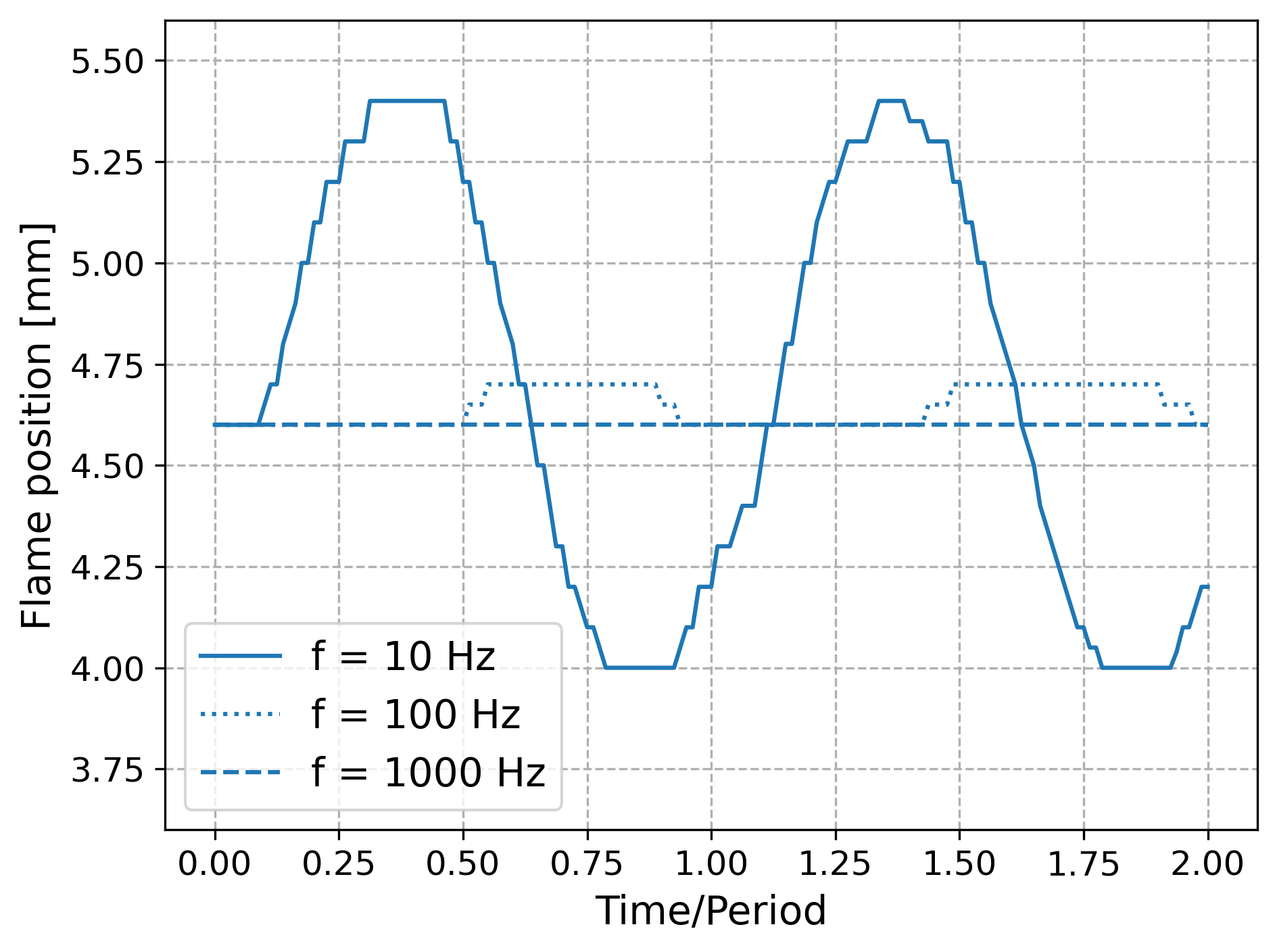}
\caption{\label{fig:Eflame-ACpos} Flame position (maximum heat release rate on central axis) under AC fields ($V_p=2\,$kV). Sub‐timestep $\Delta t_\mathrm{sub}=5\,$ns; main timestep $\Delta t_\mathrm{main}=50\,$ns.}
\end{figure}

Multi‐scale time integration is especially beneficial when resolving slow dynamic modes, such as flame response to low‐frequency AC fields. We simulate two full AC periods at 2\,kV for frequencies $f=10\,$Hz, $100\,$Hz, and $1000\,$Hz. Obviously, the computational cost is inversely correlated with the frequency. Figure~\ref{fig:Eflame-ACpos} shows flame‐position vs.\ time: at 10\,Hz, the flame oscillates sinusoidally with saturation plateaus, indicating a quasi‐steady response; at 100\,Hz, motion is minimal; at 1000\,Hz, the flame remains essentially fixed, implying suppression of the ionic‐wind effect. These trends agree with experiments \cite{park_dynamic_2018} and the 1D study in \cite{kabbaj_response_2023}.

\newpage
\section{Conclusions}
In this work, we developed and demonstrated a unified fluid model for non-thermal plasmas interacting with reacting flows, implemented in the high-performance solver \textit{reactPlasFoam}. By embedding the 0D gas–plasma kinetics solver ChemPlasKin into OpenFOAM library, we solve the full set of mass, momentum, species, enthalpy, and Poisson equations alongside on-the-fly EBE updates for the electron energy distribution. Key numerical techniques include AMR for multiscale resolution, dynamic load balancing for parallel EBE solves, and sub-cycling time integration across disparate physical modes. These methods allow accurate and efficient capture of streamer propagation, spark discharge, flame-guided plasma, and ionic-wind effects.

We benchmarked reactPlasFoam against six other codes for streamer propagation, Cantera for 1D premixed flame, and experimental data for nanosecond spark discharges. We also demonstrated, for the first time, streamer propagation inside a flame and modeled counterflow-flame response to DC and AC fields to test electrohydrodynamic force predictions. Across these cases, the solver reproduces reference results while achieving up to an order-of-magnitude speedup through our DLB and multiscale time-step schemes. These successes show that fully coupled, high-fidelity plasma–combustion simulations can now be performed at manageable computational cost, opening a new door to systematic numerical studies of two-way plasma–flame coupling. This work may also encourage more implementations of ChemPlasKin into CFD codes to achieve similar functionality.

Looking ahead, several avenues promise further improvements in the PAC modeling capabilities.  First, data-driven surrogates for the EBE could replace expensive on-the-fly solves, reducing cost while retaining accuracy.  Second, advanced stiffness-removal strategies for the reaction‐kinetics ODEs could accelerate the integration of highly stiff plasma–chemistry systems.  Third, porting critical components, such as chemical‐source evaluation and EBE solvers, to GPUs could yield additional speedups. We anticipate that these developments, together with ongoing efforts to test the model on three-dimensional geometries and incorporate detailed electrode–circuit coupling, will further expand the scope and impact of reactPlasFoam across both fundamental research and engineering applications.

\section*{Acknowledgments}
This research was funded by King Abdullah University of Science and Technology (KAUST) and granted access to the computational resources managed by KAUST Supercomputing Core Lab (KSL).

\clearpage
\section*{Appendix}
\appendix
% 1) Reset the figure counter
\setcounter{figure}{0}
% 2) (Optional) Change numbering style to “A‐1, A‐2, …”
\renewcommand{\thefigure}{A\arabic{figure}}

\section{Task Allocation Mapping Algorithm}

Several strategies are available to mitigate load imbalance when offloading tasks among parallel processors. One conventional option is the \emph{master--workers model}, a centralized approach in which a single master processor gathers global task information, redistributes tasks, and instructs all other processors. While this method is straightforward to implement and can achieve uniform workload distribution, the master processor often becomes a communication hotspot, severely hindering scalability.

A second method is the \emph{sender--receiver pairing model}, in which each overloaded processor (OP) offloads tasks to exactly one designated underloaded processor (UP). This model features simple communication and minimal message matching overhead; however, it lacks flexibility and may yield only marginal load balancing.

A third approach distributes each OP's surplus tasks evenly across all UPs, effectively creating a shared ``task pool.'' Although this strategy can improve balance beyond simple pairings, it introduces significant communication complexity. Specifically, each overloaded processor communicates with every underloaded one, resulting in $\mathcal{O}(mn)$ pairings (where $m$ is the number of OPs and $n$ is the number of UPs), which can increase the risk of MPI congestion and complex message matching.

\begin{figure}[H]
\centering
\includegraphics[width=0.5\linewidth]{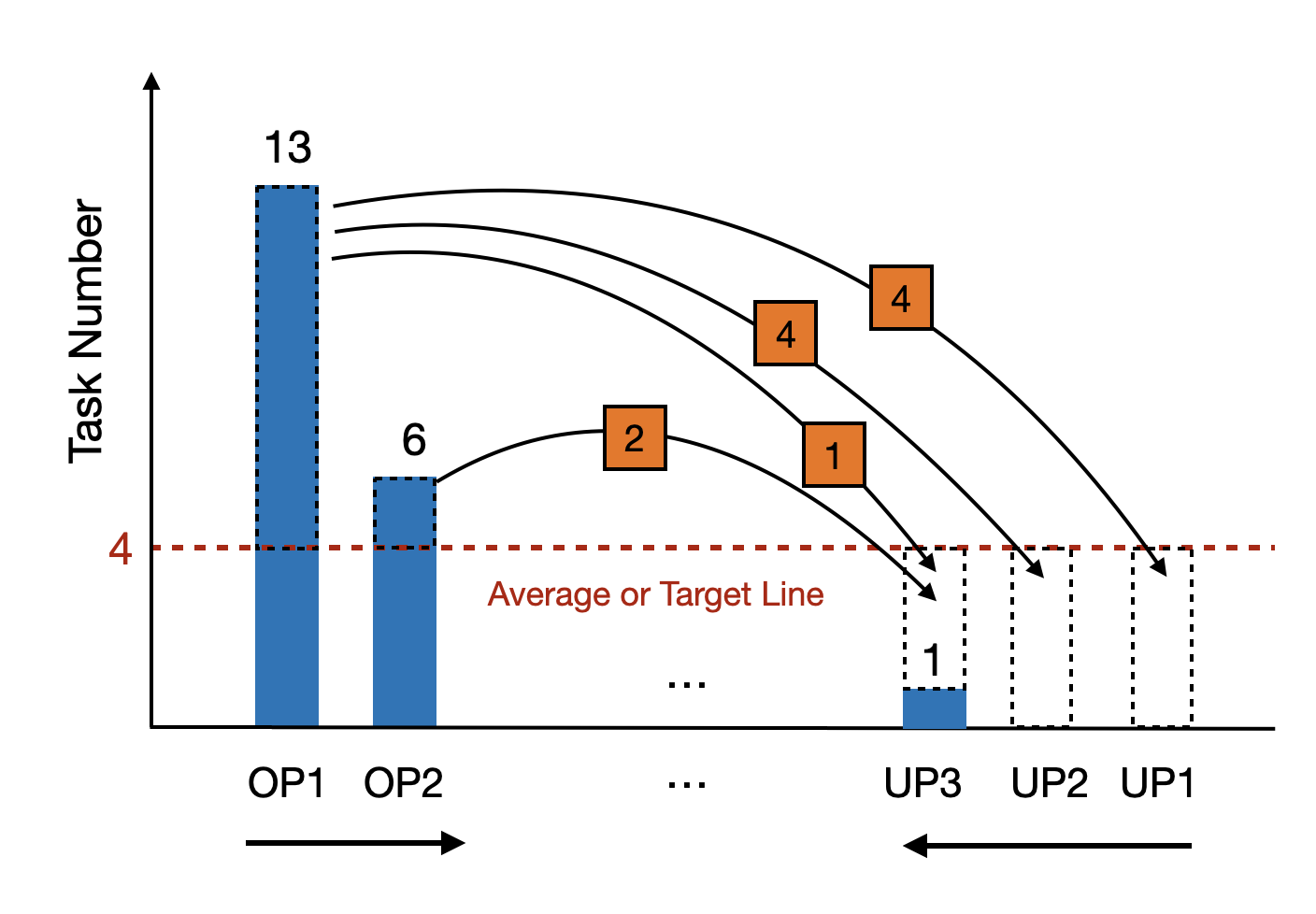}
\caption{\label{fig:task_alloc} Illustration of task allocation from Overloaded Processors (OP) to Underloaded Processors (UP). Dashed rectangles represent extra tasks on OPs or available capacity on UPs.}
\end{figure}

\paragraph{Task Allocation Mapping.} 
In our work, we found that none of these three strategies satisfied the requirements for both scalability and balanced workload distribution in all conditions. We therefore developed a \emph{task allocation mapping} algorithm that balances flexibility and communication efficiency. Figure~\ref{fig:task_alloc} provides a simplified example of how tasks are allocated. The heaviest overloaded processor (OP1), possessing nine surplus tasks, distributes four tasks each to UP1 and UP2 (filling their capacities), and then offloads the remaining task to UP3. OP2, being less heavily loaded, attempts to fill UP3 first once UP1 and UP2 can no longer receive tasks.

\begin{figure}[H]
\centering
\includegraphics[width=0.8\linewidth]{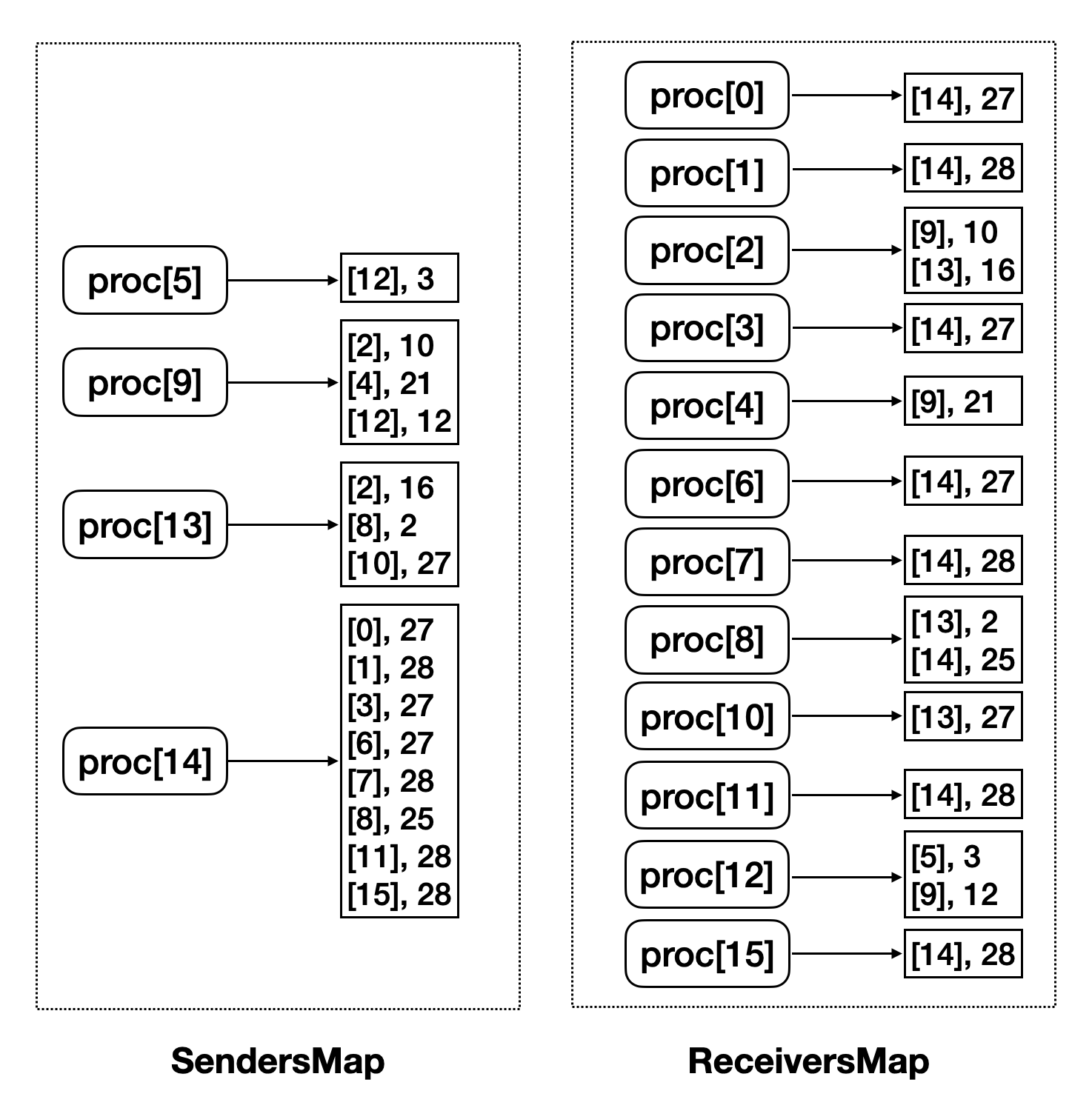}
\caption{\label{fig:DLB-maps} Sample task allocation map from both senders' and receivers' perspectives on 16 processors. Original task list: $\{1, 0, 2, 1, 7, 34, 1, 0, 1, 71, 1, 0, 13, 73, 246, 0\}$ (total of 447). Rounded average: 28.}
\end{figure}

Upon constructing the global task allocation map, each processor knows the exact senders and receivers for all tasks. For example, Figure~\ref{fig:DLB-maps} demonstrates how processor~[5] sends three tasks to processor~[12], while processor~[12] simultaneously receives 12 tasks from processor~[9]. The final map is built on every processor, ensuring consistency without a master node.

\paragraph{Load Imbalance Degree.} 
To quantify the imbalance, define
\[
    I_b \;=\; \frac{\max(L_i) - \bar{L}}{(N_p - 1)\,\bar{L}}, \quad \text{for}\; N_p > 1,
\]
where $L_i$ is the integer load of processor~$i$, $N_p$ is the total number of processors, and $\bar{L}$ is the mean load. If $\max(L_i) = \bar{L}$, then $I_b=0$ indicating perfect balance; if $\max(L_i) = N_p\,\bar{L}$, then $I_b=1$ indicating complete imbalance.

Assuming each task has the same computational cost and communication is negligible, the ideal speedup obtained by reducing $I_b$ from its initial value $I_b^0$ to $I_b^1$ is
\[
    S(N_p) \;=\; \bigl(I_b^0 - I_b^1\bigr)\, N_p.
\]

\begin{figure}[H]
\centering
\includegraphics[width=0.8\linewidth]{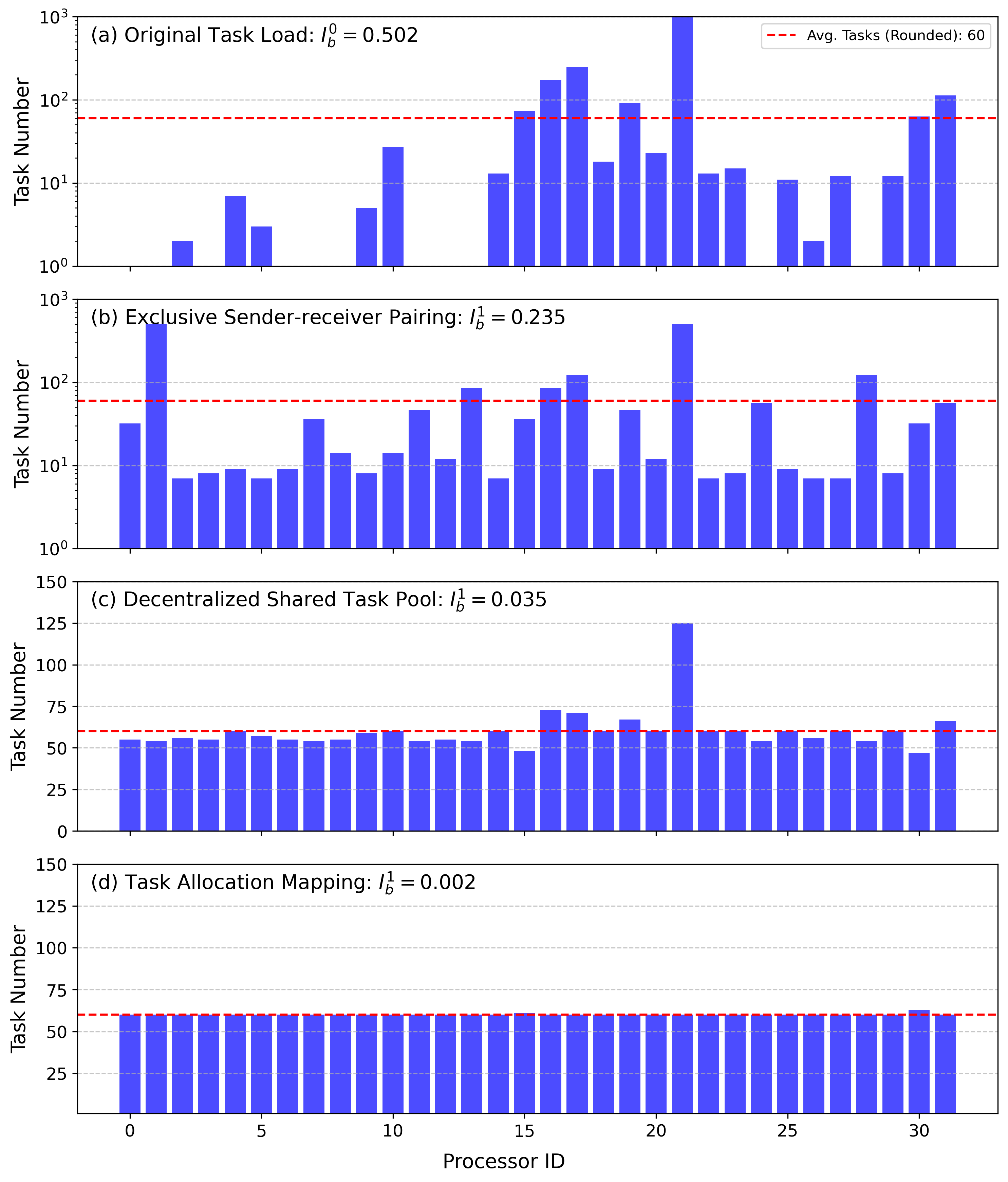}
\caption{\label{fig:DLB-method} Comparison of different load balancing strategies for a given load vector $L$.}
\end{figure}

\paragraph{Example and Speedup.} 
Consider a 32-processor system with a load distribution 
\[
   L \;=\; \{1, 0, 2, 1, 7, 3, 1, 0, 1, 5, 27, 0, 1, 0, 13, 73,\; 173, 246, 18, 92, 23, 996, 13, 15, 0, 11, 2, 12, 0, 12, 63, 113\}.
\]
Figure~\ref{fig:DLB-method} compares the original distribution against various load balancing methods. The task allocation mapping algorithm can achieve a speedup of around 16 under these conditions.

Overall, the proposed task allocation mapping algorithm effectively distributes tasks by building a global map of overloaded and underloaded processors, avoiding the centralized master bottleneck, reducing communication overhead compared to a fully shared pool, and mitigating the limitations of strict sender--receiver pairing.

\newpage
\bibliographystyle{unsrt}
\bibliography{My_Library}

\end{document}